\documentclass[pdflatex,sn-mathphys-num]{sn-jnl}
\usepackage{graphicx}
\usepackage{multirow}
\usepackage{amsmath,amssymb,amsfonts}
\usepackage{amsthm}
\usepackage{mathrsfs}
\usepackage[title]{appendix}
\usepackage{xcolor}
\usepackage{textcomp}
\usepackage{manyfoot}
\usepackage{booktabs}
\usepackage{algorithm}
\usepackage{algorithmicx}
\usepackage{algpseudocode}
\usepackage{listings}
\usepackage{siunitx}
\usepackage{braket}
\usepackage{filecontents}
\usepackage[style=nature]{biblatex}
\defbibfilter{MethodsFilter}{
  segment=1
  and not segment=0
}
\renewcommand{\thesubsection}{\arabic{subsection}}

\begin{document}
\title[Article Title]{Purcell-enhanced spin-phonon coupling with a single color-center}
\author*[1]{\fnm{Graham} \sur{Joe}}\equalcont{These authors contributed equally to this work.}\email{grahamjoe20@gmail.com}
\author[1]{\fnm{Michael} \sur{Haas}}
\equalcont{These authors contributed equally to this work.}
\author[1,2]{\fnm{Kazuhiro} \sur{Kuruma}}
\author[1]{\fnm{Chang} \sur{Jin}}
\author[3,4]{\fnm{Dongyeon Daniel}\sur{Kang}}
\author[1]{\fnm{Sophie W.} \sur{Ding}}
\author[1,5,6]{\fnm{Cleaven} \sur{Chia}}
\author[1]{\fnm{Hana} \sur{Warner}}
\author[7,8,9]{\fnm{Benjamin} \sur{Pingault}} 
\author[10]{\fnm{Bartholomeus} \sur{Machielse}} 
\author[11]{\fnm{Srujan} \sur{Meesala}}
\author*[1]{\fnm{Marko} \sur{Lončar}}\email{loncar@g.harvard.edu}
\affil[1]{\orgdiv{John A. Paulson School of Engineering and Applied Sciences}, \orgname{Harvard University}, \orgaddress{\street{9 Oxford Street}, \city{Cambridge}, \postcode{02138}, \state{Massachusetts}, \country{USA}}}
\affil[2]{\orgdiv{Current Address: Research Center for Advanced Science and Technology}, \orgname{The University of Tokyo}, \orgaddress{\street{4-6-1 Komaba}, \city{Meguro-ku}, \postcode{153–8505}, \state{Tokyo}, \country{Japan}}}
\affil[3]{\orgdiv{Center for Quantum Technology}, \orgname{Korea Institute of Science and Technology (KIST)}, \orgaddress{\street{5 Hwarang-ro 14-gil, Seongbuk-gu}, \city{Seoul}, \postcode{02792}, \country{Republic of Korea}}}
\affil[4]{\orgdiv{Division of Quantum Information}, \orgname{KIST School, Korea University of Science and Technology (UST)}, \orgaddress{\street{5 Hwarang-ro 14-gil, Seongbuk-gu}, \city{Seoul}, \postcode{02792}, \country{Republic of Korea}}}
\affil[5]{\orgdiv{Current Address: Quantum Innovation Centre (Q.InC)}, \orgname{Agency for Science Technology and Research (A*STAR)}, \orgaddress{\street{2 Fusionopolis Way, Innovis \# 08-03}, \city{Singapore}, \postcode{138634}, \country{Republic of Singapore}}}
\affil[6]{\orgdiv{Current Address: Institute of Materials Research and Engineering (IMRE)}, \orgname{Agency for Science Technology and Research (A*STAR)}, \orgaddress{\street{2 Fusionopolis Way, \#Innovis 08-03}, \city{Singapore}, \postcode{138634}, \country{Republic of Singapore}}}
\affil[7]{\orgdiv{Materials Science Division}, \orgname{Argonne National Laboratory}, \orgaddress{\street{9700 S Cass Ave}, \city{Lemont}, \postcode{60439}, \state{Illinois}, \country{USA}}}
\affil[8]{\orgdiv{Pritzker School of Molecular Engineering}, \orgname{University of Chicago}, \orgaddress{\street{5640 S Ellis Ave}, \city{Chicago}, \postcode{60637}, \state{Illinois}, \country{USA}}}
\affil[9]{\orgdiv{Q-NEXT}, \orgname{Argonne National Laboratory}, \orgaddress{\city{Lemont}, \postcode{60439}, \state{Illinois}, \country{USA}}}
\affil[10]{\orgname{IonQ, Inc.}, \orgaddress{\street{1284 Soldiers Field road}, \city{Boston}, \postcode{02135}, \state{Massachusetts}, \country{USA}}}
\affil[11]{\orgdiv{Department of Electrical and Computer Engineering, and Smalley-Curl Institute}, \orgname{Rice University}, \orgaddress{\street{
6100 Main St}, \city{Houston}, \postcode{77005}, \state{Texas}, \country{USA}}}

\begin{refsection}
\abstract{The radiative properties of emitters are inherently linked to their surrounding environment \cite{cohen-tannoudji_atom-photon_1998}. Placing an electromagnetic resonator around emitters can enhance spontaneous emission, as shown by Purcell in the 1940s \cite{purcell_spontaneous_1946}. This approach is now routinely used in quantum computing and communication to channel photons emitted by atoms into well-defined modes and control atom-photon interactions \cite{walther_cavity_2006, blais_circuit_2021, reiserer_cavity-based_2015, bienfait_controlling_2016, nguyen_integrated_2019, bhaskar_experimental_2020, uppu_quantum-dot-based_2021}. For solid-state emitters, such as color-centers, the host lattice introduces an acoustic environment, allowing excited atoms to relax by emitting phonons \cite{meesala_strain_2018, astner_solid-state_2018}. Here we observe the acoustic Purcell effect by constructing a specially engineered, microwave-frequency nanomechanical resonator around a color-center spin qubit in diamond. Using a co-localized optical mode of the structure that strongly couples to the color-center's excited state, we perform single-photon-level laser spectroscopy at milliKelvin temperatures and observe ten-fold faster spin relaxation when the spin qubit is tuned into resonance with a 12 GHz acoustic mode. Additionally, we use the color-center as an atomic-scale probe to measure the broadband phonon spectrum of the nanostructure up to a frequency of 28 GHz. Our work establishes a new regime of control for quantum defects in solids and paves the way for interconnects between atomic-scale quantum memories \cite{lemonde_phonon_2018} and qubits encoded in acoustic and superconducting devices \cite{schutz_universal_2017}.}
\maketitle

\section*{Main}
\addcontentsline{toc}{section}{Main Text}
It is well-established that an open continuum of electromagnetic modes can cause an excited atom to undergo radiative decay through spontaneous emission \cite{cohen-tannoudji_atom-photon_1998}. Careful engineering of an atom's electromagnetic environment using cavities and waveguides can be used to control its spontaneous emission rate via the well-known Purcell effect \cite{purcell_spontaneous_1946}. In cavity quantum electrodynamics (QED) and modern quantum science, this capability is harnessed at the single quantum level to control interactions between qubits and photons \cite{walther_cavity_2006}. Representative examples include Josephson junction qubits coupled to microwave resonators \cite{blais_circuit_2021}, atoms and ions in Fabry-Perot cavities \cite{reiserer_cavity-based_2015}, solid-state spins in microwave cavities \cite{bienfait_controlling_2016}, and solid-state optical emitters in nanophotonic devices \cite{nguyen_integrated_2019, bhaskar_experimental_2020, uppu_quantum-dot-based_2021}. The physics of an atom in an open continuum also applies to the emission of acoustic waves from an artificial atom in a solid. Such a process of phonon emission is particularly prominent in several color-centers \cite{meesala_strain_2018, astner_solid-state_2018} and rare-earth ion impurities \cite{orbach_spin-lattice_1997} where it limits spin relaxation times at low temperatures.

Structuring a solid and crafting an acoustic environment with well-defined modes around a color center qubit could endow the field of quantum acoustodynamics \cite{chu_quantum_2017, arrangoiz-arriola_resolving_2019} with atomic-scale, long-lived memories. This situation is analogous to cavity QED with light and microwaves. Phonons in bulk crystalline solids \cite{yang_mechanical_2024,diamandi_quantum_2024} and compact nanomechanical devices \cite{maccabe_nano-acoustic_2020,bozkurt_quantum_2023, wallucks_quantum_2020} have been shown to have long lifetimes at milliKelvin temperatures, and can transduce quantum information in various quantum systems including superconducting qubits and optomechanical devices \cite{manenti_circuit_2017,meesala_non-classical_2024,chu_perspective_2020, delaney_superconducting-qubit_2022}. These prospects have motivated experiments on color-center spin qubits in nanomechanical resonators, including motion-sensing of radio-frequency diamond nanomechanical resonators with nitrogen-vacancy centers \cite{ovartchaiyapong_dynamic_2014, meesala_enhanced_2016}, coherent control of color-center spins with resonant microwave-frequency acoustic waves \cite{maity_coherent_2020,maity_mechanical_2022,dietz_spin-acoustic_2023,whiteley_spinphonon_2019}, and control of phonon-induced orbital relaxation processes using the bandgap of phononic crystals \cite{kuruma_controlling_2025}. A central challenge in this context has been to realize a resonator-color center system that provides high spin-phonon co-operativity, with the threshold for quantum-coherent interactions being unity. In this regime, acoustic radiation from the atomic-scale spin qubit is channeled predominantly into a single acoustic mode relative to the open acoustic continuum. Here we demonstrate such a system by embedding a highly strain-sensitive color-center, the silicon vacancy (SiV) center in diamond \cite{meesala_strain_2018} in a specially engineered nanomechanical resonator.

We observe the acoustic Purcell effect between the SiV spin qubit and the 12 GHz acoustic mode of the nanomechanical resonator, near its motional ground state at millikelvin temperatures. The resonator is created in an optomechanical crystal (OMC) \cite{eichenfield_optomechanical_2009} that can co-localize photons at the optical frequency linking the spin qubit with the color-center's excited state. This design provides a minimally invasive window to probe the spin using single-photon-level laser pulses, thereby mitigating the laser-induced heating of the acoustic environment at cryogenic temperatures \cite{maccabe_nano-acoustic_2020} which limited our previous work \cite{joe_high_2024}. Simultaneously, the OMC allows us to probe the properties of the 12 GHz mechanical mode independent of its interaction with the SiV-center spin. Furthermore, we use the SiV spin qubit to measure the phonon spectrum of the nanostructure over a broad frequency range from 9 to \SI{27}{\giga\hertz}. Our results unlock new possibilities for hybrid quantum systems with color-center quantum memories and fundamental studies of coherence limits in atomic-scale quantum defects.

\subsection*{A Single Color-Center Coupled to an Optomechanical Crystal}
\addcontentsline{toc}{subsection}{A Single Color-Center Coupled to an Optomechanical Crystal}
\begin{figure}[h!]
\centering
\includegraphics[width=\textwidth]{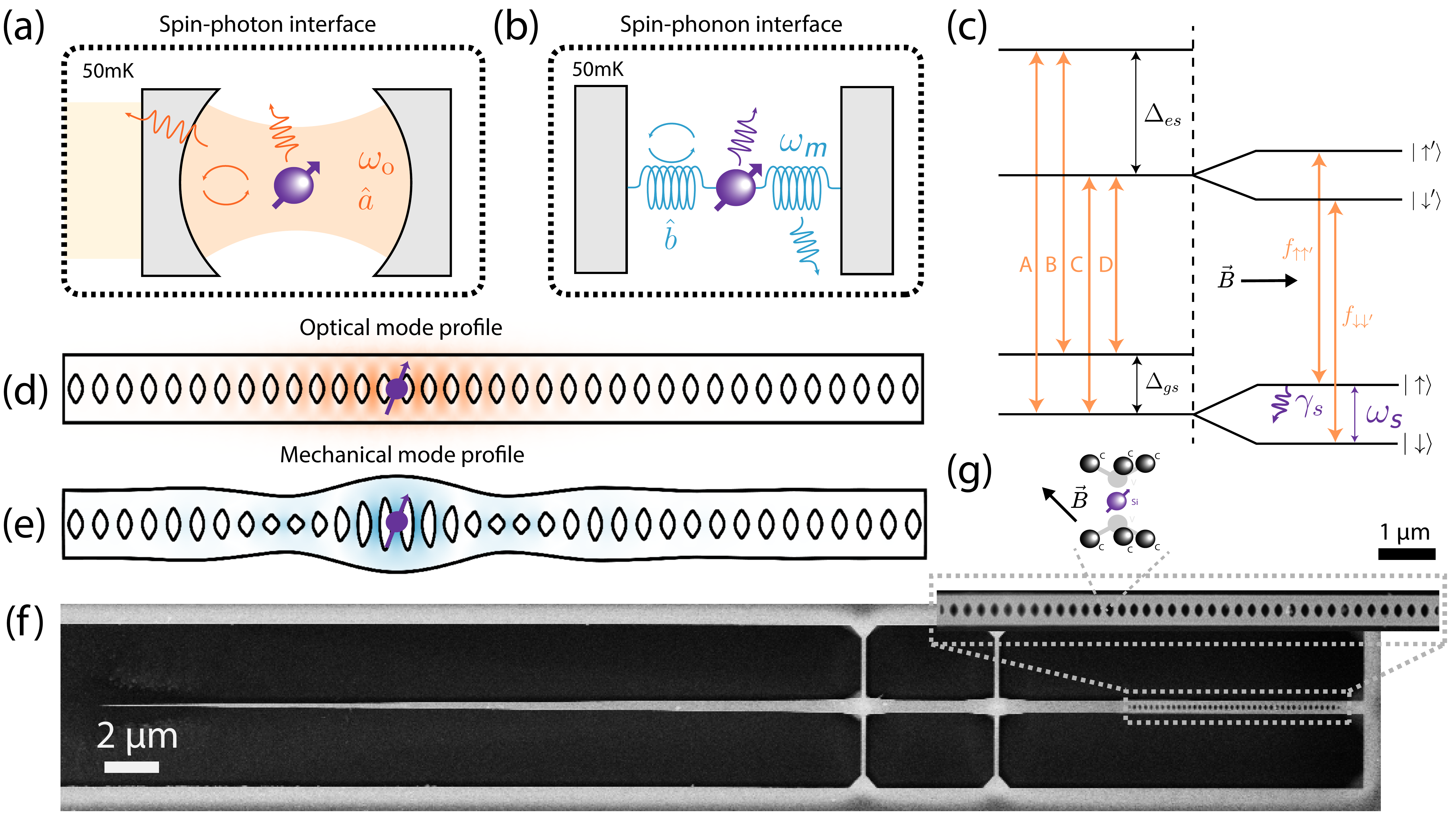}
\caption{\textbf{The coupled SiV-OMC system.} Illustration of our experimental system that simultaneously realizes efficient (a) spin-photon interactions, and (b) spin-phonon interactions. The SiV center is embedded in an optomechanical resonator which simultaneously acts as a one-sided photonic crystal cavity (orange) with a resonance frequency, $\omega_o$ and the photon annihilation operator, $\hat{a}$ and a mechanical resonator (blue) with a resonance frequency, $\omega_m$ and the phonon annihilation operator, $\hat{b}$. (c) Relevant electronic level structure of the SiV center. The splittings between orbital branches in the ground state, $\Delta_{gs}$ and the excited state, $\Delta_{es}$ and the four optical transitions at zero magnetic field A, B, C and D are indicated. The optical cavity frequency, $\omega_o$ is tuned into resonance with the C-line. An applied magnetic field \textbf{B} further splits the energy levels, defining a spin qubit on the lower orbital branch in the ground state with transition frequency $\omega_s$ which can couple to mechanical modes. Two optical transitions $f_{\uparrow\uparrow\prime}$ and $f_{\downarrow\downarrow'}$ can be used to initialize and readout the spin qubit's state. (d) Simulated electric field magnitude of the transverse electric (TE)-like optical mode and the approximate location of the implanted SiV center. (e) Simulated strain magnitude (blue) and exaggerated displacement profile of the $\sim$\SI{12}{\giga\hertz} mechanical breathing mode of the device (see SI section 2), with the approximate location of the implanted SiV center. (f) Scanning electron micrograph (SEM) of the device studied: a tapered diamond nanophotonic waveguide on the left is contacted by a tapered optical fiber to couple light in and out of the device. Two sets of tethers are created between widened sections of the diamond waveguide and the bulk diamond substrate for mechanical support and thermal contact. The optomechanical cavity on the far right is shown in detail in the zoomed in SEM in (g), with the molecular structure of the SiV center.}
\label{fig1}
\end{figure}

The system studied in this work consists of a silicon vacancy (SiV) center, whose optical C transition \cite{hepp_electronic_2014} (Fig. 1c) is coupled to a high quality factor (Q) optical mode of the surrounding OMC resonator (Fig. 1a), while its spin transition simultaneously couples to mechanical modes of the same structure (Fig. 1b). Previous work has demonstrated that an efficient and quantum-coherent spin-photon interface can be realized by integrating SiV centers with optical nanocavities \cite{nguyen_integrated_2019}. Furthermore, it has been proposed \cite{meesala_strain_2018} that an efficient and quantum-coherent spin-phonon interface could be realized by embedding SiV centers within a high-Q and small-mode-volume acoustic cavity \cite{burek_diamond_2016, joe_high_2024, li_ultracoherent_2024}. The emitter-optical and spin-mechanical cavity setups, schematically illustrated in Figs. 1a and 1b are realized simultaneously in our experimental system.

The optical resonator used is a single-sided 1D photonic crystal cavity: the mirror on the left of Fig. 1a is less reflective than the mirror on the right such that light can be coupled in and out of the cavity via an optical waveguide mode to maximize photon collection efficiency. The emitter-photon and emitter-phonon interactions are governed by Jaynes-Cummings type Hamiltonians. The optical cooperativity $C_o=\frac{4g_{so}^2}{\kappa_o\gamma_o}$ is a figure of merit for the spin-photon interface where a value $C_o>1$ indicates quantum-coherent spin-photon interactions. Here $g_{so}/(2\pi)$ is the vacuum emitter-photon coupling rate, $\omega_o/(2\pi)$ is the frequency of the optical mode, $\kappa_o/(2\pi)$ is the total loss rate of the optical mode, and $\gamma_o/(2\pi)$ is the natural linewidth of the emitter. Similarly, we can define the $T_1$- and $T_2\textsuperscript{*}$-based spin-mechanical cooperativities as $C_{T1}=\frac{4g_{sm}^2}{\kappa_m\gamma_{s,o}}$ and $C_{T2^*}=\frac{4g_{sm}^2}{\kappa_m\gamma_{s,o}^*}$ when the mode temperature is sufficiently low and thermal occupation can be neglected \cite{bozkurt_quantum_2023}. Here $g_{sm}/(2\pi)$ is the vacuum spin-phonon coupling rate, $\omega_m/(2\pi)$ is the frequency of the mechanical mode, $\kappa_m/(2\pi)$ is the mechanical dissipation rate, $\omega_s/(2\pi)$ is the spin transition frequency, $\gamma_{s,o}/(2\pi)$ is the spin relaxation rate into the continuum of the acoustic environment, and $\gamma_{s,o}^*/(2\pi)$ is the decoherence rate of the spin, including non-Markovian dephasing due to slow magnetic field fluctuations of the nuclear spin environment \cite{sukachev_silicon-vacancy_2017}.

Our devices consist of a rectangular cross-section diamond waveguide with a one-dimensional array of eye-shaped air holes designed to support a high-Q optical (Fig. 1d) mode and a high-Q mechanical (Fig. 1e) breathing mode (see Methods section \ref{fabrication} for fabrication details and Supplementary Information (SI) section 1 for design details). Light is coupled into the cavity region via a tapered fiber - diamond waveguide interface \cite{burek_fiber-coupled_2017, nguyen_integrated_2019} (Fig. 1f). We measure a bare optical cavity frequency of $\omega_o/(2\pi) = 409.7$ THz with an optical Q-factor Q$_{o} \approx 27,000$ (see SI section 3 for details). The resonant frequency of the optical mode can be tuned into resonance with the SiV zero-phonon line (ZPL) at 737 nm (406 THz) (see Methods section \ref{frequency-tuning}), giving rise to an efficient spin-photon interface. This capability enables probing the spin state using ultra-low optical probe powers on the order of \SI{1}{\pico\watt}, thus limiting the laser-induced heating of mechanical modes that had plagued previous experiments on SiV centers in optomechanical crystals \cite{joe_high_2024}. Due to the significant mode overlap between the TE-like optical mode and the $\sim$\SI{12}{\giga\hertz} mechanical breathing mode of our device, there is a significant optomechanical coupling (see SI section 2) primarily due to the photoelastic effect \cite{burek_diamond_2016}. This allows for direct probing of the properties of the mechanical mode ($\omega_m, \kappa_m$) directly using optomechanical spectroscopy \cite{eichenfield_optomechanical_2009, burek_diamond_2016, joe_high_2024}, and independently of spin-phonon coupling.

\begin{figure}[h!]
\centering
\includegraphics[width=\textwidth]{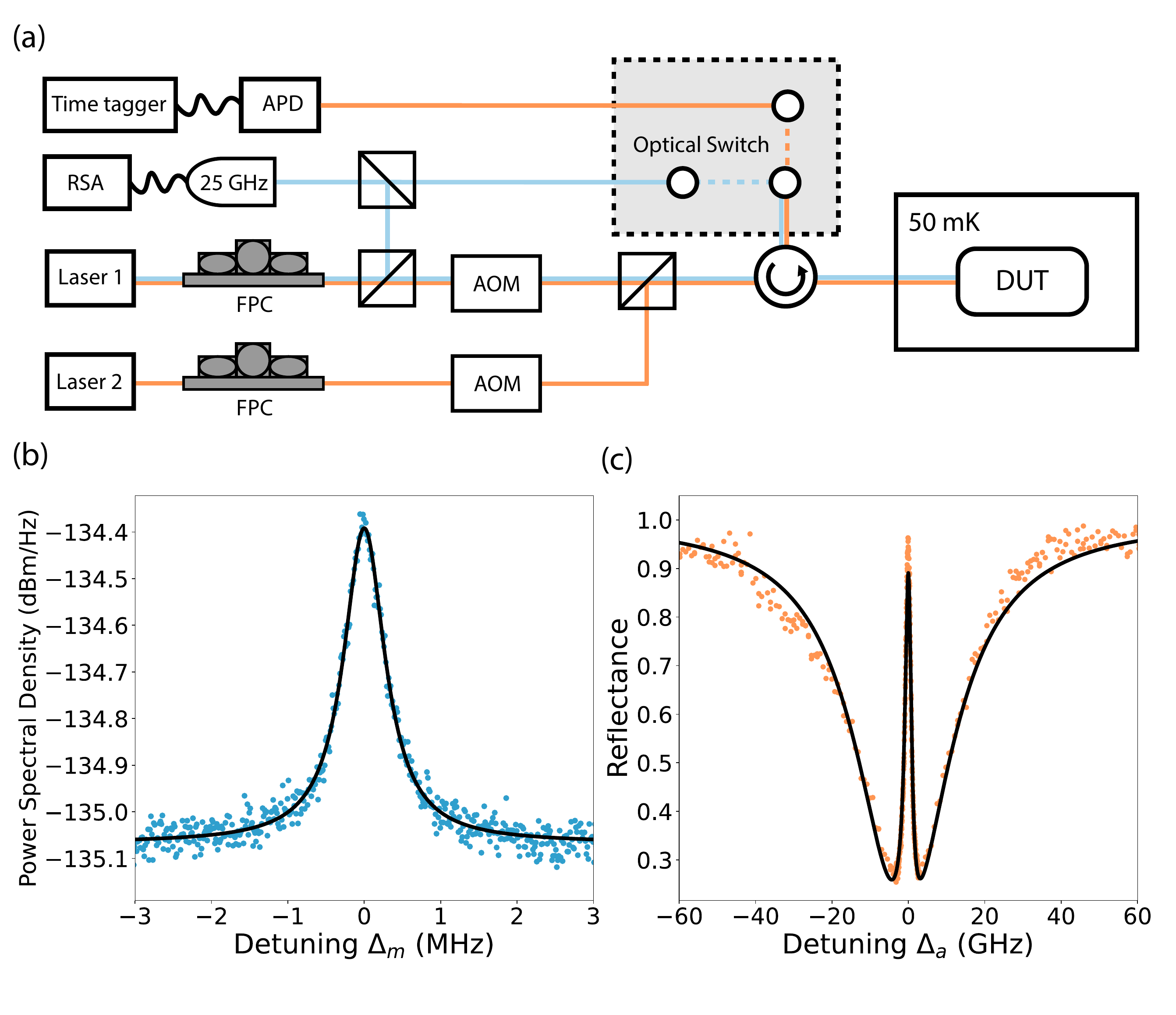}
\caption{\textbf{Characterization setup and measurements.} (a) Simplified experimental setup diagram. A tunable, continuous-wave laser (Laser 1) is used as a resonant excitation source to optically probe the device. A second laser (Laser 2) can be introduced as needed, and is combined with Laser 1 on a beamsplitter. Laser polarization states are controlled with fiber polarization controllers (FPCs) and the optical pulse sequence is generated using acousto-optic modulators (AOMs). An optical circulator is used to route the light reflected from the device under test (DUT), which is detected either by an avalanche photodiode (APD) connected to a time-tagger for photon counting (orange optical path), or by a high bandwidth photodetector (25 GHz) in a heterodyne scheme for optomechanical measurements (blue optical path) using a real-time spectrum analyzer (RSA). (b) Noise power spectral density (NPSD) of thermal motion of the 12.06 GHz mechanical breathing mode of the device measured at 4K at a laser detuning, $\Delta_o=(\kappa_{o}/2)/(2\pi)$ from the optical cavity and at an input optical power of \SI{50}{\micro\watt}, $\Delta_m = (\omega-\omega_m)/(2\pi)$ where $\omega/(2\pi)$ is the spectrum analyzer frequency. (c) Zero-magnetic-field optical reflection spectrum of the device after the gas deposition frequency tuning procedure, with the SiV C-line (transition frequency $\omega_a/(2\pi)$) on resonance with the optical cavity, $\Delta_a=(\omega_l-\omega_a)/(2\pi)$. A distinct peak in the reflection spectrum can be observed due to hybridization of the optical mode with the SiV C-line. A fit to the optical spectrum allows us to infer the cavity QED parameters of our spin-photon interface: $(g_{so}, \kappa_o, \gamma_o)/(2\pi) = (3.6, 15, 0.11)$ GHz.}
\label{fig2}
\end{figure}

A simplified diagram of the optical setup used to measure our devices is depicted in Fig. 2a (see Methods section \ref{full-setup} for full setup details). We study our devices in a dilution refrigerator setup equipped with a vector magnet, at a base temperature of $\sim$\SI{44}{\milli\kelvin} under experimental conditions. We first probe the mechanical breathing mode of the structure via the optomechanical coupling. These measurements are performed at zero magnetic field with the optical transitions of the SiV center far detuned from the optical cavity. Therefore, the SiV does not participate in the optomechanical interaction. The probe laser (frequency $\omega_l$) is detuned by $\Delta_o = (\omega_l - \omega_o) / (2\pi) = \pm (\kappa_{o}/2)/(2\pi)$ from the cavity to operate at the maximum optomechanical susceptibility of a sideband-unresolved system \cite{aspelmeyer_cavity_2014}. The optically detected noise power spectral density (NPSD) of thermal motion of the mechanical breathing mode at 4K, recorded from a real-time spectrum analyzer (RSA), is shown in Fig. 2b. We observe a breathing mode resonance frequency $\omega_m/(2\pi) = 12.06$ GHz close to the design value, with intrinsic mechanical linewidth $\kappa_m/(2\pi)\approx350$ kHz (intrinsic mechanical quality factor Q$_m \approx$ 34,000 measured at 4K, see SI section 4).

To compensate for the frequency difference between the SiV-center's optical transition and the optical cavity mode, which is caused by finite fabrication tolerances (see SI section 1), we investigated two methods for cavity frequency tuning: in situ gas deposition frequency tuning (gas tuning), and atomic layer deposition (ALD) of alumina with minimal additional gas deposition (see Methods section \ref{frequency-tuning}). Although the former is commonly used in cryogenic nanophotonic cavity QED experiments \cite{nguyen_integrated_2019}, it was observed to reduce the Q-factor of the mechanical resonator upon material deposition. The latter technique allowed us to minimize this effect. The optical Q-factor is not affected by either tuning method. The optical reflection spectrum of the cavity-SiV system operated on resonance with an input power of $\sim$\SI{1}{\pico\watt} ($n_c\sim 2\times10^{-4}$) is shown in Fig. 2c, only a single SiV center is observed to couple to the optical mode. The spectrum features a sharp peak within the cavity dip - a signature of the hybridization between the atomic and optical degrees of freedom \cite{waks_dipole_2006, sipahigil_integrated_2016}. A fit to this spectrum and data at different atom-cavity detunings yields $(g_{so}, \kappa_o, \gamma_o)/(2\pi) = (3.6, 15, 0.11)$ GHz corresponding to an emitter-photon cooperativity of $C_o=31$. Note that in the very low intracavity photon number regime ($n_{c}\ll1$) used for all experiments involving the SiV center, the optomechanical photon-phonon interaction is negligible.

\subsection*{Observation of the Acoustic Purcell Effect}
\addcontentsline{toc}{subsection}{Observation of the Acoustic Purcell Effect}
\begin{figure}[H]
\centering
\includegraphics[width=\textwidth]{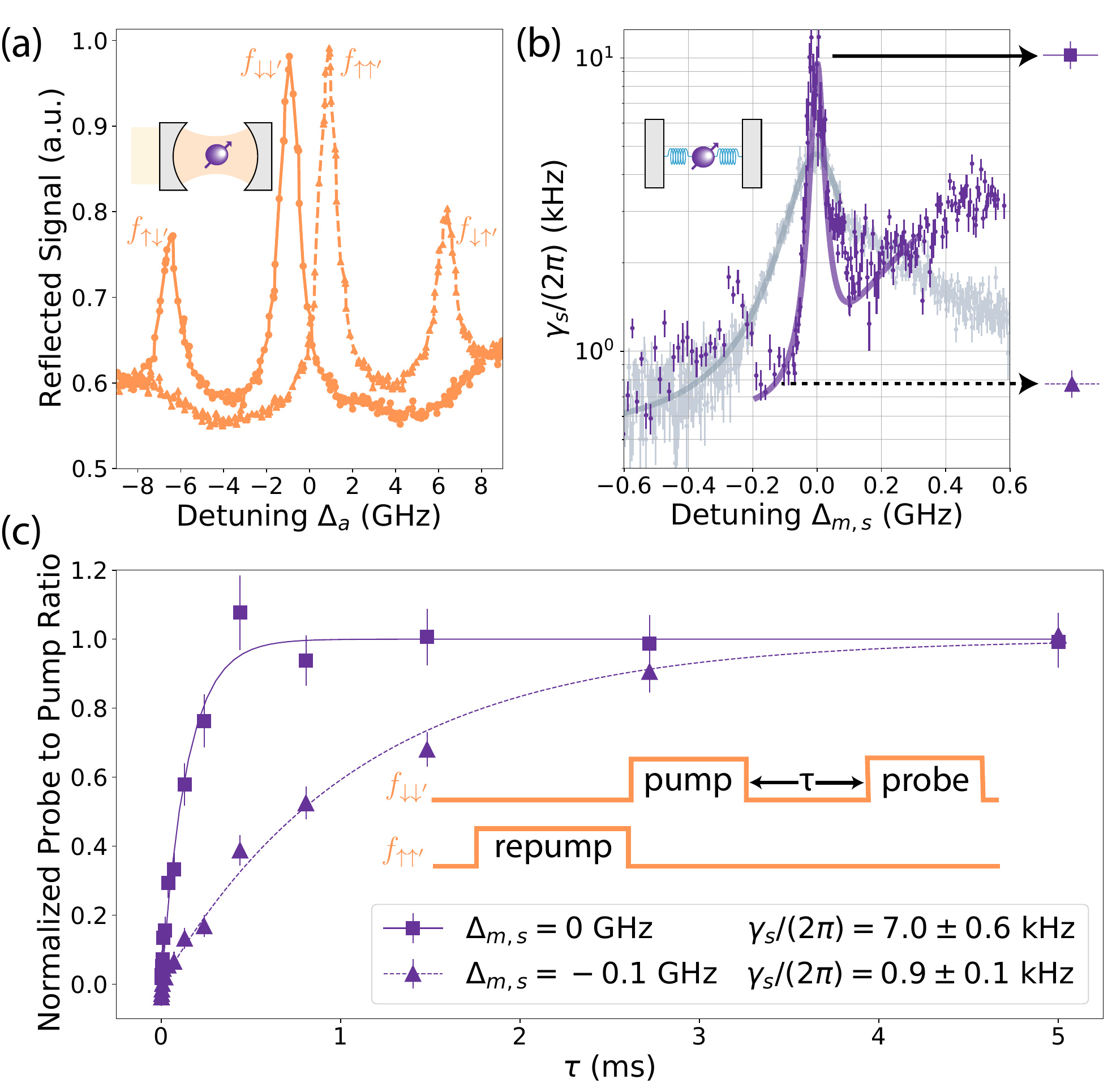}
\caption{\textbf{Observation of the acoustic Purcell effect.} (a) Continuous-wave optical reflection spectrum of the spin-photon interface (inset) measured around the SiV center's C-line after the application of a 2.5 kG magnetic field oriented $\theta\approx$ \SI{55}{\degree} relative to the SiV center's high-symmetry axis, differences between the optical line frequencies measure the spin transition frequency $\omega_s$ (see SI section 5). A repump laser of equal power is resonant with the $f_{\uparrow\uparrow\prime}$ ($f_{\downarrow\downarrow\prime}$) transition while measuring the spectrum shown with the solid (dashed) trace. (b) Resonance in the spin decay rate measured when the spin qubit frequency, $\omega_s/(2\pi)$ is tuned across the $\sim$\SI{12}{\giga\hertz} breathing mode due to the spin-phonon interaction (inset). The data is taken with a $\theta\approx$ \SI{55}{\degree} magnetic field in two separate runs where the optical cavity is tuned into resonance with SiV center's optical transition using the gas tuning method (gray data points, fit is performed to a truncated data set excluding the non-Lorentzian shape of the right-hand side of the peak), and a combination of ALD and gas tuning (purple data points, fit is performed to a truncated data set and uses a heuristic piecewise linear model of the background), resulting in a fit resonance linewidth of $\sim$\SI{200}{\mega\hertz} and  $\sim$\SI{35}{\mega\hertz}, respectively. The error bars represent the standard deviation of the spin decay rate. (c) Spin population decay curves for the ALD+gas-tuned device with the spin transition frequency $\omega_s/(2\pi)$ at a detuning $\Delta_{m,s}$ from the $\sim$\SI{12}{\giga\hertz} mechanical mode of \SI{0}{\giga\hertz} (solid line, corresponding to the peak in panel b) and -\SI{0.1}{\giga\hertz} (dashed line, corresponding to the baseline in panel b). Error bars represent the standard deviation of the normalized probe to pump ratio. (inset) Optical pulse sequence used to measure the SiV center's spin decay rate is shown in the inset of the plot (see SI section 6 for details of the pulse sequence).}
\label{fig3}
\end{figure}

To define a spin qubit, we lift the spin degeneracy by applying a magnetic field of 2.5 kG along the z- axis in the lab frame ($\theta\approx $ \SI{55}{\degree} with respect to the SiV center's high symmetry axis, see Methods section \ref{angular-dependence}) this results in a spin transition frequency, $\omega_s/(2\pi)$, which can be measured via resonant laser excitation spectroscopy of the Zeeman split optical transitions (see SI section 5). We initialize the SiV spin state by optically pumping on the Zeeman-split transitions using the repump and pump pulses shown in the inset of Fig. 3c. At millikelvin temperatures, after initializing the spin into the $\ket{\uparrow}$ state, the population decay rate $\gamma_s/(2\pi)$ is proportional to the acoustic density of states of the surrounding structure at $\omega_s/(2\pi)$. For the range of magnetic fields and the system temperature in these measurements, the spin population is expected to be predominantly in the $\ket{\downarrow}$ state in the steady state, since $k_BT \ll\hbar\omega_s$. In the SI section 7, we use spin relaxation measurements to infer that the SiV center's environment is at a temperature of $\sim 150$ mK. At this temperature, the acoustic modes in the environment are close to their quantum ground state, and the $\sim12$ GHz breathing mode is expected to have a thermal occupancy of 0.02 quanta.

We studied the SiV center’s spin population decay rate, $\gamma_s/(2\pi)$ as a function of the spin transition frequency, $\omega_s/(2\pi)$ which is tuned by varying the magnitude of the applied magnetic field. At step changes in the magnetic field, we perform optical spectroscopy to track the optical transition frequencies f$_{\uparrow\uparrow\prime}$ and f$_{\downarrow\downarrow\prime}$ (see SI section 5). When the detuning between the spin transition frequency and the mechanical breathing mode frequency $\Delta_{m,s} = (\omega_s - \omega_m)/(2\pi)$ approaches zero, a distinct peak in the spin population decay rate is observed, as shown in Fig. 3b, indicating a Purcell enhancement of the spin relaxation rate. The gray data points depict data observed when the optical cavity frequency tuning is performed in situ using the gas deposition method to shift the optical resonance from 409.7 to 406.7 THz (see Methods section \ref{frequency-tuning}). This approach results in a broad resonance in the Purcell-enhanced spin relaxation rate with a linewidth of $\sim 200$ MHz. On the other hand, the purple data points are obtained after conformal deposition of $\sim 5$ nm of alumina using ALD to shift the optical resonance from 409.7 to 406.9 THz, followed by in situ gas tuning to shift the optical resonance to 406.7 THz. This approach results in a linewidth of $\sim 35$ MHz for the Purcell enhancement. In the former case, the measured linewidth is a factor of $\sim$500 larger than the intrinsic mechanical linewidth of 350 kHz measured via optomechanics (see SI section 4). The deposition of the ALD alumina layer broadens the intrinsic mechanical linewidth to 650 kHz (see SI section 4). In our experimental setup, such an independent optomechanical spectroscopy measurement cannot be performed on a gas-tuned device because the requisite optical powers ($\sim$\SI{1}{\micro\watt}) boil off the deposited gas. Nonetheless, this combination of spin-relaxation and optomechanical measurements indicates that the amount of gas deposited on the device is strongly correlated with the linewidth of the Purcell enhancement due to the mechanical resonator (Fig. 3b). Since the solid deposited on the structure due to gas deposition is expected to have significantly lower speed of sound relative to diamond \cite{pederson1998problems}, we expect gas deposition to increase the acoustic radiative loss from the mechanical resonator to the bulk substrate, introducing excess damping.

For the remainder of the discussion, we present data measured on the device using the ALD cavity-tuning method and minimal residual gas deposition, with corresponding data for the gas-tuning method shown in the Methods section \ref{frequency-tuning}. Fig. 3c shows the spin population decay curves measured at zero detuning from the breathing mode (solid curve) and with a detuning of -\SI{0.1}{\giga\hertz}, clearly illustrating the Purcell enhancement in spin relaxation between the on and off-resonant cases. The optical pulse sequence used to measure the SiV center’s spin population decay rate at a given spin transition frequency $\omega_s/(2\pi)$ is shown in the inset (see SI section 6 for details). The data fits well to an exponential decay, allowing us to extract the spin relaxation rate. The Purcell-enhanced spin-decay rate due to a near ground state mechanical resonance where $\kappa_m\gg\gamma_s$ is given by $\gamma_s = \gamma_{s,o} + g_{sm}^2\frac{\kappa_m}{\kappa_m^2/4 + (\omega_s-\omega_m)^2}$ where $\gamma_{s,o}$ is the spin relaxation due to the environment of the spin mechanical resonator system. We measure $\gamma_s/(2\pi)\approx 1$ kHz at a detuning of -0.1 GHz from the resonance which we take as the value of $\gamma_{s,o}/(2\pi)$, and we measure $\gamma_s/(2\pi)\approx 10$ kHz on resonance. With the linewidth of the resonance measured as $\kappa_m/(2\pi)\approx35$ MHz, we infer a spin-phonon coupling rate of $g_{sm}/(2\pi)\approx 300$ kHz. Our simulations suggest that under ideal conditions of low static strain and accurate SiV placement at the breathing mode strain maximum, $g_{sm}/(2\pi)$ could attain a maximum value of 9 MHz (see SI section 8). The difference between this maximum value and the experimentally measured value is well explained by the non-ideal conditions in the studied device, including the presence of static strain (see SI section 9) and the typical expected error in SiV placement. Our measurements correspond to a $T_1$-based cooperativity $C_{T1}\approx10$. Assuming the previously measured value for the SiV center spin linewidth, $\gamma_{s,o}^*/(2\pi)\approx1$ \SI{}{\mega\hertz} \cite{nguyen_integrated_2019} in photonic crystal cavities with similar dimensions, this corresponds to a $T_2$\textsuperscript{*}-based cooperativity $C_{T2^*}\approx0.01$.

\begin{figure}[h!]
\centering
\includegraphics[width=\textwidth]{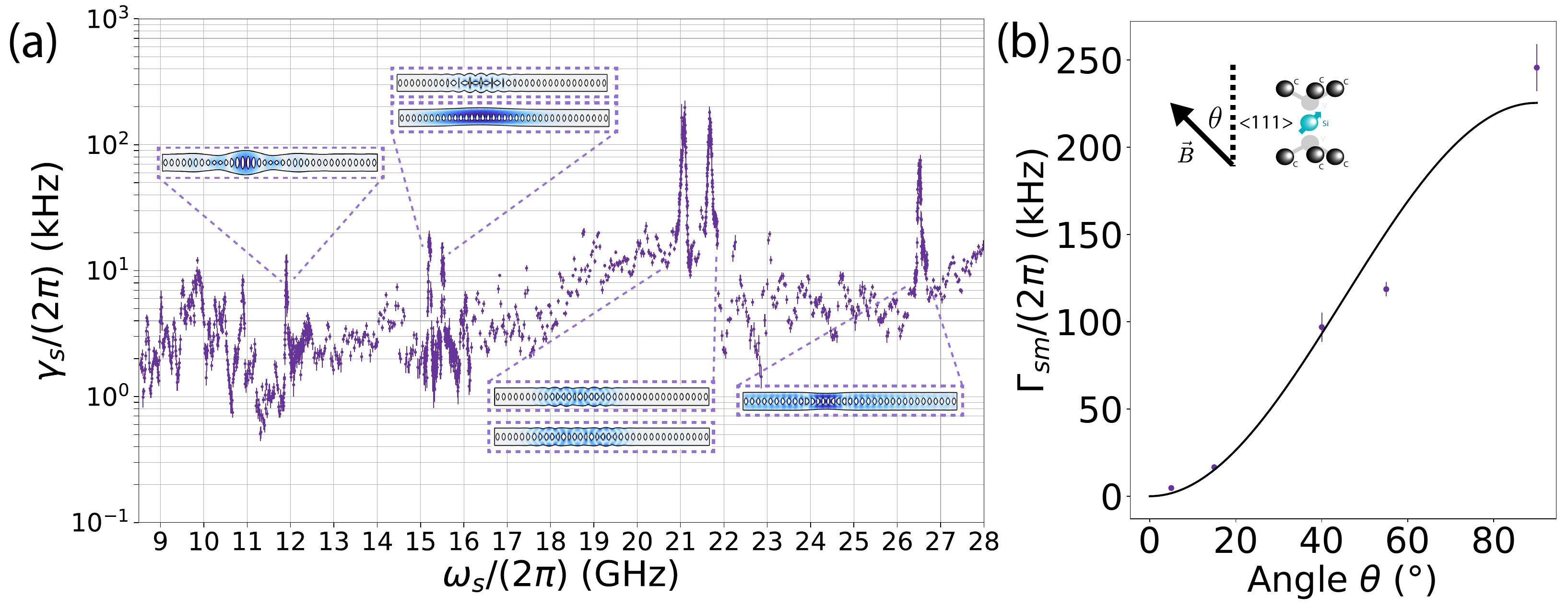}
\caption{\textbf{Broadband and angular dependence of spin decay.} (a) Broadband spectrum of the spin decay rate, $\gamma_s$ for a $\theta\approx$ \SI{55}{\degree} magnetic field. The error bars represent the standard deviation of the spin decay rate. Several resonances in addition to the $\sim$\SI{12}{\giga\hertz} breathing mode are observed and attributed to other mechanical modes of the device (insets), as confirmed by numerical simulations (see SI section 10). (b) Dependence of the spin decay rate on the angle of the applied magnetic field relative to the SiV center's symmetry axis measured on the mechanical resonance at  $\sim$21 GHz. The error bars represent the standard deviation of the $\sim$21 GHz peak amplitude. The solid line is a single-parameter fit to the relation, $\Gamma_{sm} = A \mathrm{sin}^2\theta$, the theoretical dependence expected from the SiV spin-strain interaction Hamiltonian \cite{meesala_strain_2018}.}
\label{fig4}
\end{figure}

In addition to the resonance induced by the mechanical breathing mode, several distinct features are observed in the spin decay spectrum of Fig. 4a. This broadband spectrum is taken with a $\theta\approx$ \SI{55}{\degree} aligned magnetic field. We attribute these resonances in the spin decay rate to other localized mechanical modes of the nanostructure with appreciable spin-phonon coupling to the SiV. Although these modes cannot be observed independently using optomechanical spectroscopy (see SI Section 2), numerical simulations of the acoustic mode structure of our device (using the geometric parameters extracted from SEM analysis in the SI section 1) reveal several acoustic modes at other frequencies with appreciable spin-phonon coupling. The simulated spin decay spectrum is qualitatively similar to the experimentally measured spectrum, but is also strongly dependent on the precise position of the SiV center within the OMC device (see SI section 10).

To further confirm the acoustic origin of the Purcell enhancement, we measure the dependence of the spin relaxation rate on the orientation of the magnetic field for the resonance at $\sim$\SI{21}{\giga\hertz}. Based on theoretical predictions using the SiV strain interaction Hamiltonian \cite{meesala_strain_2018}, the magnitude of Purcell-enhanced spin decay rate ($\Gamma_{sm} = \gamma_{s, \omega_m}-\gamma_{s,o}$) should increase as the magnetic field vector rotates away from the SiV center's high-symmetry axis towards the plane perpendicular to that axis due to increased mixing between the spin-orbit eigenstates (see Methods section \ref{angular-dependence} for the theoretical details and the measurement details). In Fig. 4b, the fit to the data indicates good agreement with the $\mathrm{sin}^2\theta$ dependence expected from theory.

\subsection*{Outlook}
\addcontentsline{toc}{subsection}{Outlook}
In conclusion, we observed the acoustic Purcell effect between a single color-center spin qubit and a nanomechanical resonator, and leveraged it to perform broadband phonon spectroscopy of the nanostructure. At the 12-GHz acoustic resonance of the device, we observe a $T_1$-based spin-phonon cooperativity, $C_{T1}\approx10$ and a $T_2$\textsuperscript{*}- based cooperativity, $C_{T2^*}\approx0.01$. While this is the highest spin-phonon cooperativity observed to date, we expect future advances in our experimental platform to enable $C_{T2^*}>1$, the regime for most applications involving quantum-coherent interactions. In the present work, the cooperativity is primarily limited by excess mechanical damping (increased $\kappa_m$) from the material deposition (gas deposition or ALD alumina with minimal gas deposition) required to tune the co-localized optical mode into resonance with the SiV. An increase in cooperativity of at least two orders of magnitude is feasible with tuning methods that avoid material deposition, such as electrostatically-tunable cavities \cite{frank_programmable_2010}. In this regime, the system can be used to perform bi-directional transduction of quantum states between the spin qubit and the acoustic mode. The acoustic resonator in our platform is amenable to co-design with recently developed electro-optomechanical \cite{bozkurt_quantum_2023} and piezo-optomechanical quantum transducers \cite{meesala_non-classical_2024, jiang_efficient_2020}, strengthening prospects for interconnecting color-center spin-based quantum memories with superconducting qubits\cite{neuman_phononic_2021}. Such a hybrid quantum system would combine the exceptional coherence and atomic-scale footprint of color-center spin qubits with fast, high-fidelity superconducting qubit gates, and provide unique opportunities for next-generation quantum networks \cite{covey_quantum_2023} and distributed quantum information processing \cite{monroe_large-scale_2014}. 

Beyond applications involving hybrid quantum hardware, we expect the spin-phonon cooperativity can be further boosted by reducing acoustic radiative loss into the continuum through phononic bandgap shielding, previously used to achieve ultra-high-Q GHz-frequency, nanoscale acoustic resonators in silicon \cite{maccabe_nano-acoustic_2020}. In this setting, our system will enable the realization of acoustic metamaterials with strong atom-phonon interactions and chip-scale quantum networks \cite{schutz_universal_2017, lemonde_phonon_2018} as well as fundamental studies of material noise with phonon-engineered two-level-system defects \cite{chen_phonon_2024, maksymowych_frequency_2025, yuksel_strong_2025}.

\printbibliography[segment=0]
\section*{Methods}
\addcontentsline{toc}{section}{Methods}
\newrefsegment
\subsection{Fabrication of Rectangular Cross-Section SiV-OMC Devices}\label{fabrication}
\setcounter{figure}{0}
\renewcommand{\thefigure}{Extended Data \arabic{figure}}
\begin{figure}[h!]
\centering
\includegraphics[width=\textwidth]{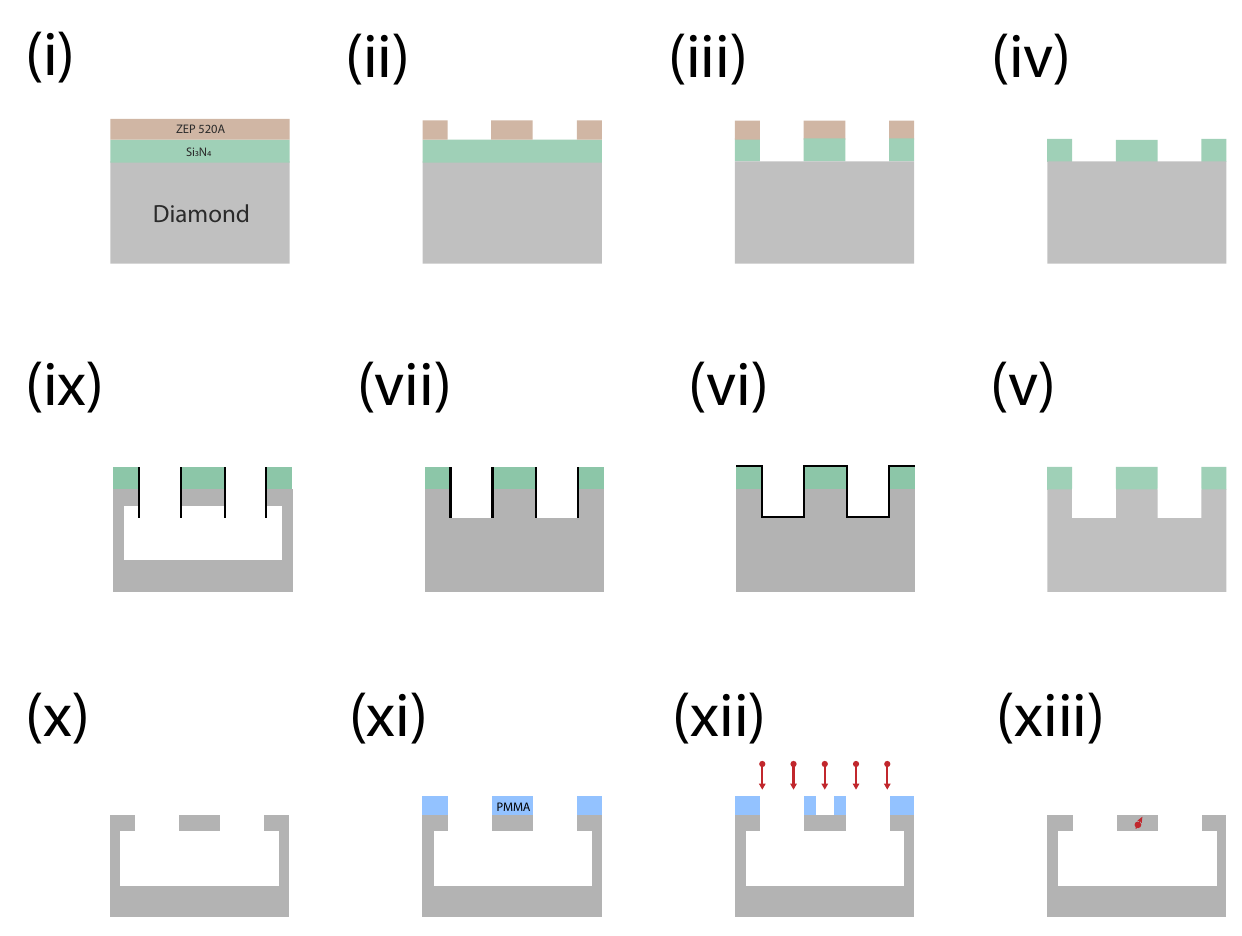}
\caption{\textbf{Fabrication procedure.} Schematic of process flow for the SiV-OMC device fabrication. The OMC devices are fabricated in diamond based on a quasi-isotropic etching technique (i-x). The SiV centers are deterministically generated and incorporated into the fabricated OMC devices by masked ion implantation process (x-xiii).  This procedure typically results in a few SiV centers per cavity, the device studied had a single detectable SiV center.}
\label{fig:fabrication_process}
\end{figure}

The rectangular cross-section SiV-OMC devices are fabricated by quasi-isotropic etching. \cite{khanaliloo_high-qv_2015, kuruma_coupling_2021, joe_high_2024} An overview of the fabrication process is shown in ED Fig. 1. We use plasma-enhanced chemical vapor deposition to make a 100 nm-thick SiN layer on the bulk diamond before spin-coating a 400 nm-thick electron beam (EB) resist (ZEP 520A) [(ED Fig. 1(i))]. The OMC pattern is created on the EB resist by EB lithography and resist development [(ED Fig. 1(ii))].  The SiN layer is etched by inductively coupled plasma-reactive ion etching (ICP-RIE) with sulfur hexafluoride (SF\textsubscript{6}) and octafluorocyclobutane (C\textsubscript{4}F\textsubscript{8}) gases [(ED Fig. 1(iii))]. After the EB resist is removed[(ED Fig. 1(iv))], the OMC pattern is transferred into the diamond by oxygen plasma-based ICP-RIE[(ED Fig. 1(v))]. We deposit  20 nm-thick Al$_2$O$_3$  by atomic layer deposition (ALD) for conformal coverage of the sample[(ED Fig. 1(vi))]. We etch the Al$_2$O$_3$  layer by RIE with Ar and Cl gases, maintaining the sidewalls of the OMC devices covered with Al$_2$O$_3$ [(ED Fig. 1(vii))]. We realize the free-standing structures using the quasi-isotropic etching technique based on an oxygen-based RIE[(ED Fig. 1(ix))]. We remove the Al$_2$O$_3$  and SiN layers by hydrofluoric acid. 

To implant SiV centers in the fabricated OMC devices, we employ a deterministic ion implantation technique \cite{nguyen_integrated_2019, evans_photon-mediated_2018}. The details of the technique can be found in our previous report \cite{machielse_quantum_2019}. Briefly, we spin-coat a PMMA layer on the fabricated devices [(ED Fig. 1(xi))] and create the implantation mask with apertures (62.5 nm x 62.5 nm) by a combination of EB lithography and resist development. The silicon ions (Si\textsuperscript{+}) are implanted only in the apertures with an energy of 150 keV. The resulting mean ion range was $\sim$75 nm from the surface, as simulated by the software package Stopping and Range of Ions in Matter. We choose a low ion implantation dose (1.0 x $10^{12}$ ions per cm$^{2}$) intended to produce roughly 3 SiV centers per aperture on average. After masked ion implantation [(ED Fig. 1(xii))], we remove the PMMA layer to complete the fabrication of OMC devices [(ED Fig. 1(xiii))]. The SiV centers in OMC devices are generated by annealing. We note that our fabricated devices typically have a few SiV centers, we use an OMC device with a single SiV center.

\subsection{Measurement Setup}\label{full-setup}
\begin{figure}[h!]
\centering
\includegraphics[width=\textwidth]{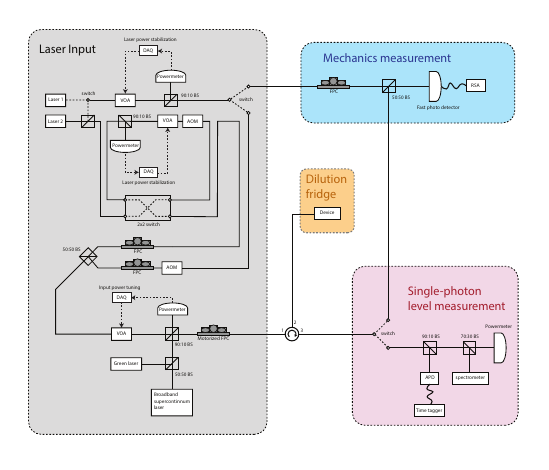}
\caption{\textbf{Detailed measurement setup.} Schematic of measurement setup. VOA: Variable Optical Attenuator, BS: Beamsplitter, DAQ: Data Acquisition Device, AOM: Acousto-Optic Modulator, FPC: Fiber Polarization Controller, APD: Avalanche Photodetector, RSA: Real-Time Spectrum Analyzer.}
\label{fig:setup}
\end{figure}

Our measurement setup is designed to enable both SiV spectroscopy and optomechanical measurements in slightly different configurations. SiV measurements require low input laser powers on the order of \SI{1}{\pico\watt} at the device, while optomechanical measurements require input powers at the device on the order of \SI{1}{\micro\watt} or more.

The primary laser used in our experiments for resonant optical excitation of SiVs is a continuous-wave Ti:Sapphire laser (MSquared, Laser 2 in ED Fig. 2). SiV measurements sometimes require a repump laser (Toptica DL Pro, Laser 1 in ED Fig. 2). Optomechanical measurements use power split off from Laser 2 as a reference local oscillator (LO) in a heterodyne scheme. We use the upper branch of the laser preparation path for optical power measurement and control. An optical switch (Photonwares) is used to direct the laser towards the sample for SiV measurements and towards a high bandwidth photodetector (Thorlabs, RXM25DF) for use as an LO in the heterodyne scheme for optomechanical measurements.

On both the upper and lower branches of the resonant laser input path, we use a voltage-controlled variable optical attenuator (Thorlabs VOA, 0-5 V) followed by a 90:10 fiber beamsplitter with a powermeter (Thorlabs) on the 10\% port for setting and locking the input power, if desired. An analog output voltage from the powermeter is measured with a data acquisition device (DAQ, National Instruments NI-USB 6363) and the input power can be locked with a PID loop. Prior to combination of the resonant laser paths on a 50:50 beamsplitter, both lasers pass through a fiber acousto-optic modulator (AOM, AA Optoelectronics MT200-R18-Fio-PM-J1-A-VSF) and a manual fiber polarization controller (FPC) which can set the relative polarization of the lasers. Each AOM is driven by a 200 MHz source which is in turn controlled by a microwave switch followed by a microwave amplifier. The microwave switch is controlled by a delay generator (Stanford Research Systems DG645). After being combined, the resonant lasers pass through a second voltage controlled VOA. Combined with the first VOA, this allows for the required $\sim$ 60 dB of dynamic range in power control for locking of the input power while achieving picoWatt-level powers at the device. On Laser 2's path, we add a 2x2 optical switch which can bypass the VOA and AOM. This is used to reduce optical path loss during optomechanical heterodyne measurements. In this case, the VOA and power meter further downstream can be used to control the power of Laser 2.

The resonant input lasers are finally combined with a green diode laser (Thorlabs DJ532-10) and a broadband super-continuum laser (NKT Photonics SuperK EXTREME) on a 2x2 fiber beamsplitter. The green laser is used for charge state repumping. The broadband laser is used for low-resolution reflection measurements in combination with a spectrometer (Princeton Instruments, HRS-750) and for fine frequency adjustment of gas-tuned OMC devices. The unused port of the 2x2 beamsplitter can be used for input power measurements. A motorized fiber polarization controller is used for automated control of the input laser polarization state during long measurements.

A fiber optic circulator customized for operation at 737 nm (Ascentta Optics) is used just before the input to the dilution fridge (Bluefors LD250) to separate the device reflection from the input with a low insertion loss of about 1 dB per pass. The optical fiber going to the device is routed through a blank KF flange with a small hole drilled through the middle, which is then sealed shut with Torr Seal. The fiber runs down the length of the fridge, and is looped and taped to the mixing plate in an attempt to thermalize the fiber to the mixing plate as well as possible before reaching the device. The fiber is then spliced to a tapered optical fiber mounted on a stack of three Attocube stages (ANPxz101/LT/HV), used for positioning the fiber relative to the sample. The base temperature of our fridge under experimental conditions is $\sim44$ mK. 

Reflection from the device can then be directed either to single-photon-level detection paths where we perform optical characterization of our devices and all SiV-related measurements, or to an optomechanical measurement path. The main detector used in for single-photon level measurements is an APD (Perkin Elmer) connected to a Time-Tagger (Swabian). We use a powermeter (Thorlabs, PM101A) for higher power measurements, and a spectrometer (Princeton Instruments, HRS-750) for low-resolution reflection measurements (5 GHz resolution, allows for live reflection spectrum monitoring e.g. during gas tuning) and fluorescence measurements.

For optomechanical measurements, device reflection is combined with an LO derived from the original laser on a 2x2 fiber 50:50 beamsplitter. The polarization of the LO is matched to the device reflection using an FPC. One output port of the 2x2 beamsplitter is not used while the other goes to a high-bandwidth silicon photodetector (Thorlabs, RXM25DF)) connected to a real-time spectrum analyzer (RSA, Tektronix 5126A). Modulation of the laser reflected from the device due to mechanical modes with significant optomechanical coupling generates optical side-bands, which beat with the LO and generate a microwave frequency voltage signal that is detected on the RSA.

\subsection{Gas Tuning and ALD Cladding}\label{frequency-tuning}
\begin{figure}[h!]
\centering
\includegraphics[width=\textwidth]{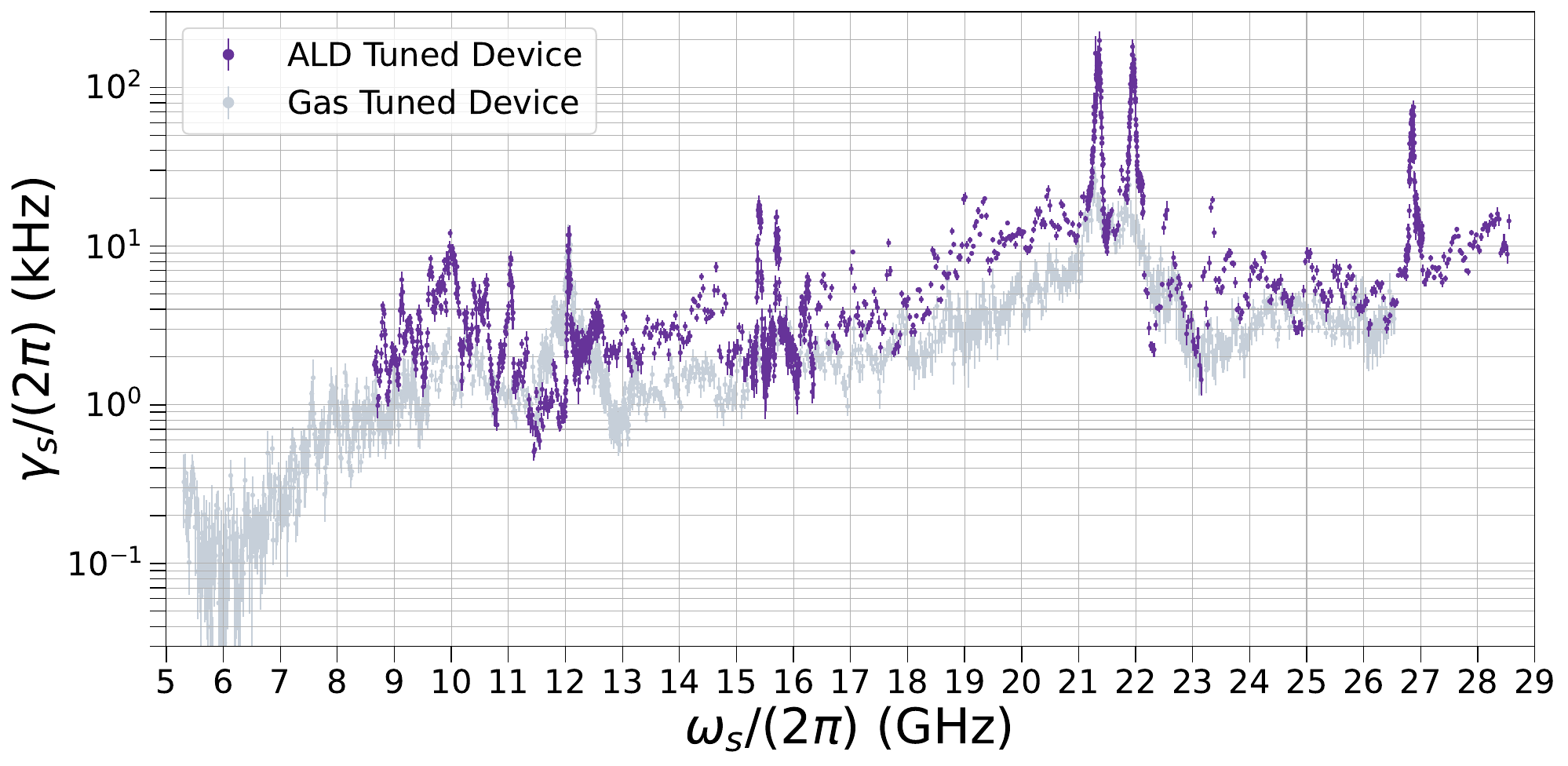}
\caption{\textbf{Broadband spin decay spectra comparison.} Comparison between the broadband spin decay spectra when the optical mode of the device is tuned entirely via gas deposition with a $\theta\approx$ \SI{90}{\degree} B-field orientation (gray) and mostly via ALD alumina cladding with minimal gas deposition with a $\theta\approx$ \SI{55}{\degree} B-field orientation (purple). The error bars represent the standard deviation of the spin decay rate.}
\label{fig:gas_tuning_spectra}
\end{figure}

The optical mode of our device can be tuned into resonance with the SiV center’s C-line by in-situ gas deposition. Gas is delivered to the sample through a copper tube thermally connected to the 4K plate of the dilution refrigerator. This tube inside the fridge is mated to a stainless steel tube which runs outside of the fridge. Resistive heaters and dedicated thermometers are installed on the gas tube, allowing control of the tube temperature. During the cooldown, after the fridge reaches a temperature of about 4K, the tube is heated above the boiling temperature of inert gases, which can then deposit on the sample. We observed that once the gas tube reaches a temperature in the range of 35-45K at the heater (around the 4K stage), the optical resonance of our device begins to red-shift, indicating gas deposition. The material deposited is likely Nitrogen, the component of air with the lowest sublimation temperature at high vacuum ($\sim$22 K \cite{satorre_density_2008}, likely the temperature of the gas tube near the sample). Using this method, we achieve a coarse wavelength shift of the cavity resonance to a longer wavelength with respect to the SiV C line. After this step, a fine-tuning step is performed to blue-shift the cavity resonance onto the SiV C line. This is achieved via controlled boiling of the deposited gas with a high-power broadband laser (NKT Photonics, Super K supercontinuum white light laser) through the tapered fiber interface, which blue-shifts the resonance. During this tuning process, the optical resonance frequency is monitored in real time on a spectrometer (Princeton Instruments HRS-750). FDTD simulations of our structure yield a frequency tuning rate of 65 GHz/nm of deposited gas, so the thickness of the deposited gas layer used to tune the bare cavity into resonance with the SiV center's C-line is $\sim$45 nm. 

To further study the effect of gas deposition, we also studied these devices with minimal gas deposition by cladding the devices in $\sim5$nm of alumina by atomic layer deposition (ALD). Following deposition, sample mounting and cool-down, the optical mode of this device was \SI{736.8}{\nano\meter} (\SI{406.825}{\tera\hertz}), and the intrinsic mechanical linewidth broadened to \SI{650}{\kilo\hertz}. Note that thermorefractive shifts and potential residual deposition of material on the device during cool-down can slightly shift the resonance frequency relative to the ambient room temperature case. Further gas deposition was required to achieve resonance with the SiV center’s C-line, but a comparatively small amount relative to the bare device. The gray trace in ED Fig. 3 corresponds with the case where the optical cavity was tuned entirely by gas deposition and the magnetic field alignment of $\theta\approx90$\SI{}{\degree}, while the purple trace corresponds to the case of ALD alumina cladding with minimal gas deposition and a magnetic field alignment of $\theta\approx55$\SI{}{\degree}. However, as shown in the angle-dependent measurements in ED Fig. 4, the angle of the applied magnetic field has a negligible effect on the linewidth of the Purcell enhancement due to an acoustic resonance. In both spectra, a distinct peak is observed around 12 GHz where the mechanical breathing mode is expected. The linewidth of this feature is substantially broadened in the gas-tuned device relative to the ALD-tuned device. In addition to the $\sim$\SI{12}{\giga\hertz} feature shown in Fig. 3a, several other features in the spin decay spectrum were observed to narrow in linewidth with some features becoming visible that were previously not apparent in the fully gas tuned spectrum. We attribute these features to other mechanical modes of the structure, which are not observed in the optomechanically transduced noise power spectrum measurements because of insufficient optomechanical coupling (see SI section 2).

\subsection{Angle-Dependent Spin Decay Spectra}\label{angular-dependence}
\begin{figure}[h!]
\centering
\includegraphics[width=\textwidth]{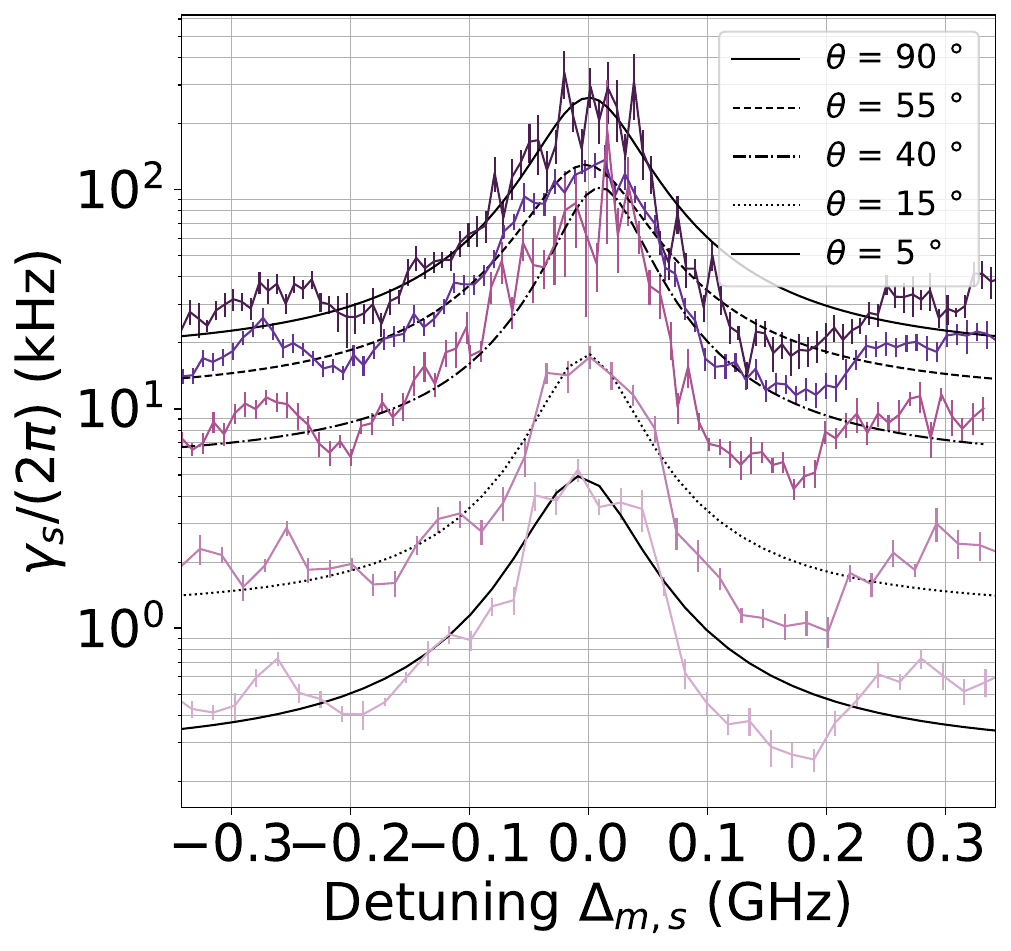}
\caption{\textbf{Angular dependence of spin decay spectra.} Spin decay rate spectra and fits at different magnetic field orientations relative to the axis of the SiV center for an example feature around 21 GHz. The error bars represent the standard deviation of the spin decay rate.}
\label{fig:angle_dependence}
\end{figure}

Most of the measurements in this work used a magnetic field angle of \SI{55}{\degree} relative to the high symmetry axis of the SiV center. This orientation corresponds to the laboratory z-axis, allowing measurements to be performed using a single coil of the superconducting vector magnet for experimental simplicity. Although a magnetic field aligned at \SI{90}{\degree} would maximize the SiV spin decay rate due to strain susceptibility (see the results of Fig. 4b in the main text), the \SI{55}{\degree} configuration still provides a substantial perpendicular field component ($B_{\perp}$), making it sufficient for most measurements presented in this manuscript.

\begin{figure}[h!]
\centering
\includegraphics[width=\textwidth]{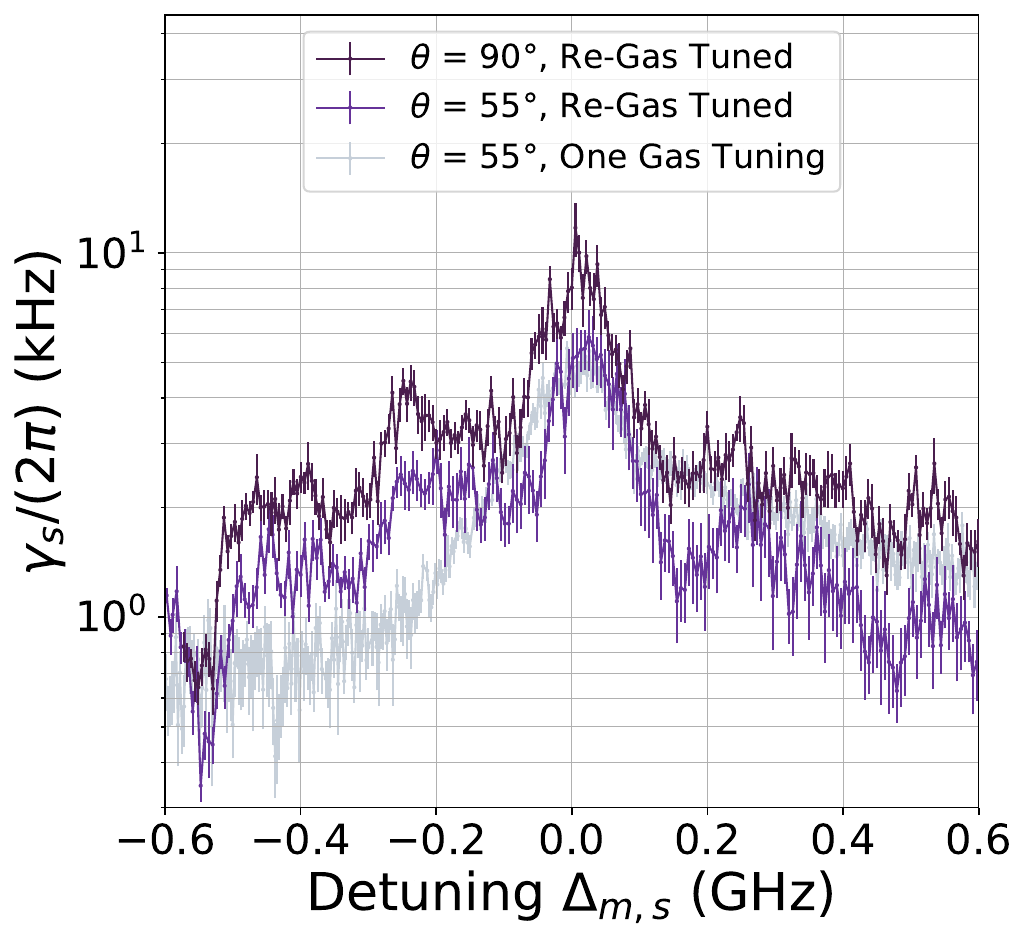}
\caption{\textbf{Breathing mode study.} Spin decay spectra for the 12 GHz mechanical breathing mode under different gas-tuning-only experimental conditions. The gray curve is the data presented in Fig. 3b with a 55\SI{}{\degree} aligned magnetic field after a single gas tuning. The lighter purple uses the same 55\SI{}{\degree} aligned magnetic field but after boiling off the deposited gas with a high power laser and re-gas tuning. The darkest purple trace is under the same deposited gas conditions as the lighter purple trace, but with a 90\SI{}{\degree} aligned magnetic field. The error bars represent the standard deviation of the spin decay rate.}
\label{fig:angle_dependence_12GHz}
\end{figure}

In Fig. 4b we studied the effect of the magnetic field orientation on the spin decay rate spectrum of an example feature around 21 GHz, the details of the spectra are shown here in ED Fig. 4. We chose to study this high frequency feature because at alignments close to the SiV axis the splitting between $f_{\uparrow\uparrow\prime}$ and $f_{\downarrow\downarrow'}$ is small, making it much easier to study high frequency modes. In ED Fig. 4 we sweep the spin transition frequency across this $\sim 21$ GHz feature by adjusting the magnetic field magnitude for five different magnetic field orientations. The amplitude and baseline of the $\sim 21$ GHz feature increase as the field alignment rotates from close to parallel with the SiV axis towards the perpendicular. Following the standard cavity QED framework for the Purcell effect \cite{cohen-tannoudji_atom-photon_1998}, we expect $\Gamma_{sm} \propto g_{sm}^2$. Further, the Hamiltonian of the SiV center dictates that the acoustic dipole moment (strain susceptibility) for the spin qubit, and hence $g_{sm}$ is proportional to proportional to the component of magnetic field perpendicular to the SiV-center's high-symmetry axis \cite{meesala_strain_2018}. In Fig. 4b, the fit to the data indicates good agreement with this $\mathrm{sin}^2\theta$ dependence expected from theory. 

In ED Fig. 5 we show measurements of the 12 GHz mode taken under gas-tuning-only conditions for both a 55\SI{}{\degree} and a 90\SI{}{\degree} aligned field. The gray curve is the data presented in Fig. 3b after a single gas tuning for a 55\SI{}{\degree} aligned magnetic field. The other two measurements were performed by then boiling off the gas with a high power laser and re-gas tuning. This repeated procedure further degraded the mechanical resonance - likely due to residual gas outside the cavity center - resulting in a mildly distorted resonance shape. The mode quality was later restored after cleaning the chip and depositing ALD Alumina. The data clearly shows the theoretically expected angular dependence of the 12 GHz mode; a more perpendicular magnetic field leads to faster spin-decay rates. A full angular dependence for the comparatively low frequency 12 GHz mode is experimentally challenging due to a crossing of the optical transitions used for spin initialization at alignments close to the SiV center's high symmetry axis, and was not performed.

\printbibliography[heading=subbibliography,filter=MethodsFilter,title={Methods References}]

\section*{Acknowledgments}
\addcontentsline{toc}{section}{Acknowledgments}
The authors would like to thank M. Bhaskar, D. Assumpcao, and C. Knaut for fruitful discussions. This research was supported by the National Science Foundation under grant number DMR-1231319, titled ``Center for Integrated Quantum Systems (CIQM)," the Army Research Office/Department of the Army under award number W911NF1810432, titled ``Ab-Initio Solid-State Quantum Materials: Design, Production, and Characterization at the Atomic Scale," the Office of Naval Research under award number N00014-20-1-2425, titled ``Quantum Information Processing With Phonons," the National Science Foundation under award number EEC-1941583, titled ``NSF Engineering Research Center for Quantum Networks (CQN)," the Air Force Office of Scientific Research under award number FA9550-23-1-0333, titled ``Quantum Phononics to Advance Quantum Information Processing," the National Research Foundation funded by the Korean government (No. RS-2022-NR068818), and Amazon Web Services under award number A50791, titled ``Partnership for Quantum Networking." G. J was supported in part by the Natural Sciences and Research Council of Canada (NSERC). K.K. acknowledges financial support from JSPS Overseas Research Fellowships (Project No. 202160592). This work was partly supported by the KIST institutional program (26E0001, 26E0011) funded by the Korea Institute of Science and Technology, and the National Research Foundation of Korea (NRF) grant (No. RS-2025-25445839), the Institute for Information and Communication Technology Planning and Evaluation (IITP) grant (No. RS-2025-25464657), and the National Research Council of Science and Technology (NST) grant (No. GTL25011-000), funded by the Korean government (MSIT). C. C was supported in part by Singapore’s Agency for Science, Technology and Research (A*STAR). H. W acknowledges financial support from the NSF GRFP fellowship. B.P. acknowledges financial support from the U.S. Department of Energy, Office of Science, Basic Energy Sciences, Materials Sciences and Engineering Division through Argonne National Laboratory under Contract No. DE-AC02-06 CH11357. B.M. is involved in developing diamond photonic technology at IonQ, Inc. This work was performed in part at the Center for Nanoscale Systems (CNS), a member of the National Nanotechnology Infrastructure Network (NNIN), which is supported by the National Science Foundation award ECS-0335765. CNS is part of Harvard University.

\section*{Author Contributions Statement}
\addcontentsline{toc}{section}{Author Contributions Statement}
G.J., M.H., B.P., S.M. and M.L. conceived the experiment. G.J. and M.H. planned the experiment. M.H. and C.C. designed and simulated the device. K.K. fabricated the devices with help from S.W.D. and B.M.. G.J, M.H. and H.W. built and maintained the cryogenic measurement system. G.J., M.H., C.J., and D.D.K performed the experiment. G.J., M.H., C.J., D.D.K., B.P., and S.M. analyzed the data and interpreted the results. All authors discussed the results and provided critical feedback. G.J., M.H., C.J., B.P. and S.M. wrote the manuscript. M.L. supervised the work.

\section*{Competing Interests}
\addcontentsline{toc}{section}{Competing Interests}
The authors claim no financial or non-financial competing interest.

\section*{Data Availability}
\addcontentsline{toc}{section}{Data Availability}
The data that support the findings in this study are available from the corresponding authors on request.
\end{refsection}
\end{document}


\title[Article Title]{Supplementary Information for ``Purcell-enhanced spin-phonon coupling with a single color-center"}
\maketitle

\begin{refsection}
\section{Unit Cell Design and SEM Analysis} \label{sem-analysis}
The goal of the OMC cavity design is to achieve an optical/mechanical cavity mode pair which simultaneously enables optical readout of the SiV center's spin properties and optomechanical readout of the mechanical mode properties. This requires high values of $\{g_{so},\ g_{om},\ g_{sm}\}$ and low values of $\{\kappa_o, \kappa_m\}$. There has been a great deal of previous work on designing mode pairs which satisfy these criteria \cite{chan_optimized_2012}. We choose the fundamental TE-like optical mode, mechanical ‘breathing mode’ pair, which features a near-optimal photoelastic overlap integral between the optical $E_y$ field component with the strain $\epsilon_{YY}$ field component, which  enables large optomechanical coupling. The TE-like optical mode also supports efficient spin-photon interactions and the mechanical breathing mode supports efficient spin-phonon interactions for an SiV oriented \SI{45}{\degree} relative to the beam. 

\begin{figure}[h!]
    \centering
    \includegraphics[width=.9\linewidth]{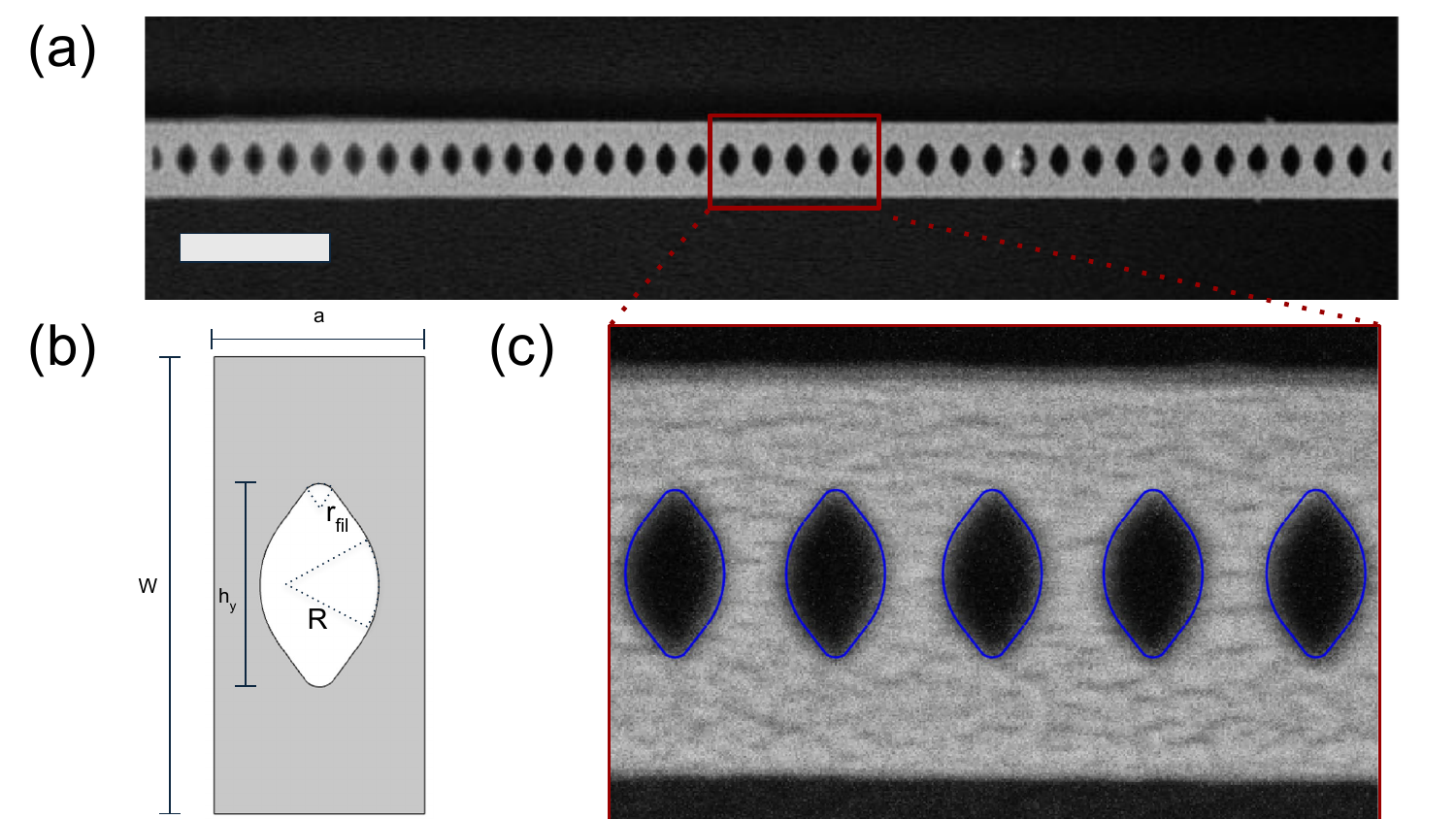}
    \caption{(a) SEM of the cavity region of the device studied. (b) Parameterized diagram of the OMC cavity unit cell. (c) SEM of 5 unit cells in the OMC cavity with best fit unit cells drawn in blue.}
    \label{fig:sem_analysis}
\end{figure}

The cavity region of the fabricated device is shown in Fig. \ref{fig:sem_analysis}. A unit cell design akin to the conventional elliptical hole shape is used (where the hole is defined by its major/minor axis lengths) except an additional parameter $R$ is introduced, which determines the radius of curvature near the hole edge.

To accurately compare the experimental results to simulation, we took detailed SEM images of each unit cell hole in the OMC. Image processing on the SEM images then extracts the fabricated unit cell parameters, with the addition of the parameter $r_{\text{fil}}$ to account for rounding at the vertical extents of the unit cell hole. For the device under study, we observe the measured parameters in Table \ref{tab:sem_vs_nominal}.

\begin{table}[h!]
    \centering
    \begin{tabular}{|c|c|c|}
        \hline
        Parameter & Measured Value & Designed Value \\ \hline
        Beam thickness ($th$) & 150 nm & 160 nm \\ \hline
        Beam width ($w$) & 530 nm & 540 nm \\ \hline
        Radius ($R$) & 110 nm & 100 nm \\ \hline
        Angle ($\theta$) & $36^\circ$ & $36^\circ$ \\ \hline
        Hole height ($h_y$) & 240 nm & 236 nm \\ \hline
        Fillet radius ($r_\text{fil}$) & 30 nm & 0 nm \\ \hline
    \end{tabular}
    \caption{Comparison of SEM and Nominal Parameter Values}
    \label{tab:sem_vs_nominal}
\end{table}

These parameters deviate from the nominal parameters of our device within normal fabrication tolerances, but do lead to measurable shifts in device properties.

\section{Simulation of the Optomechanical Cavity}\label{device-design}
\begin{figure}[h!]
    \centering
\includegraphics[width=\linewidth]{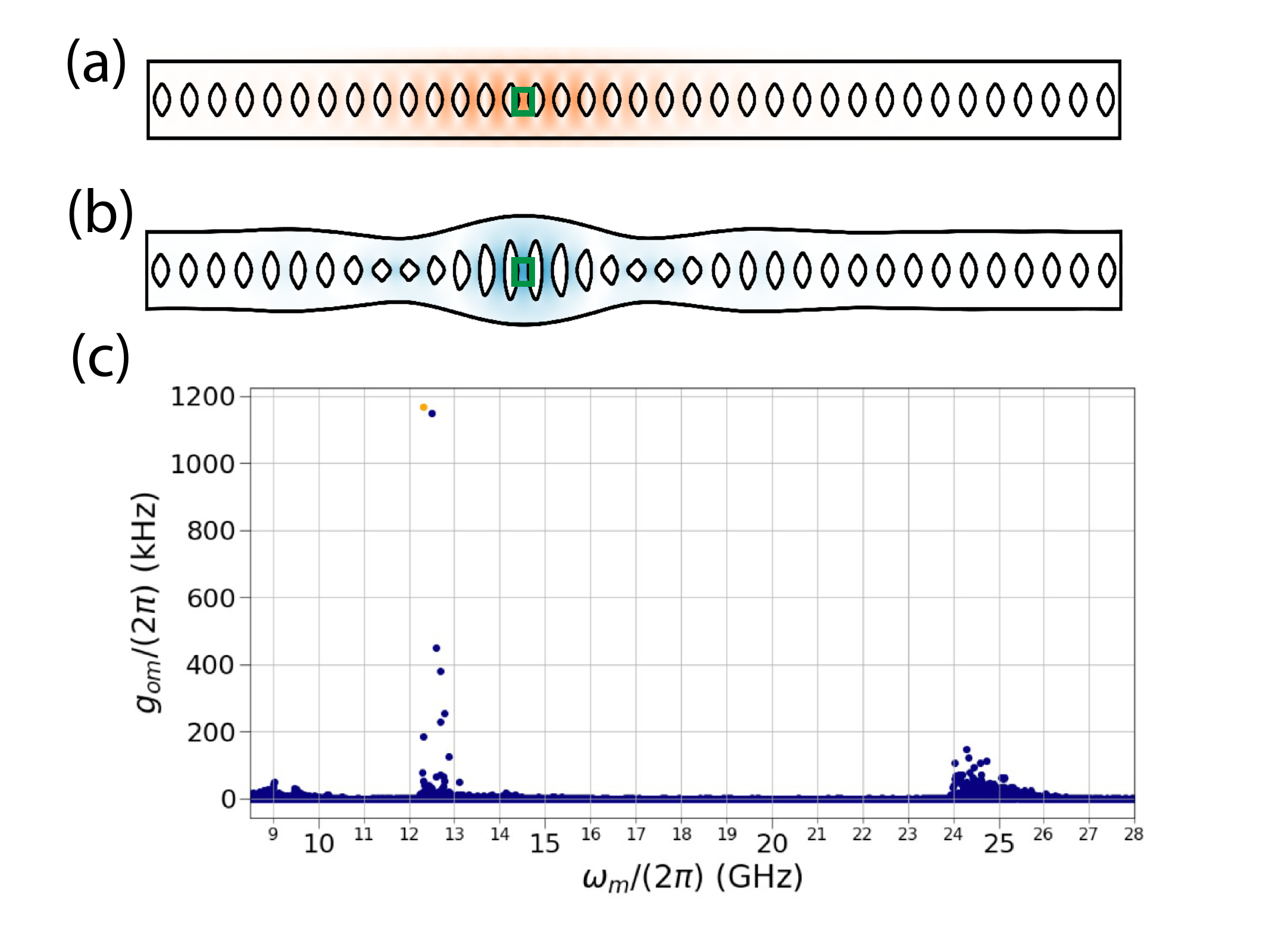}
    \caption{(a) Electric field profile of TE-like optical mode. Green box indicates masked implantation aperture, ensuring good alignment between E-field and SiV location which enables large spin-photon coupling. (b) Strain ($\epsilon_{XX} + \epsilon_{YY} + \epsilon_{ZZ}$) profile of mechanical breathing mode. (c) Calculation of optomechanical coupling rate for full spectrum of mechanical eigenmodes. The orange dot signifies the breathing mode, showing that it satisfies the large $g_{OM}$ requirement for optomechanical readout of the mechanical mode properties.}
    \label{fig:cavity_gOM}
\end{figure}

The optical TE-like and mechanical breathing modes were simulated using the geometric parameters obtained from SEM images (see SI section \ref{sem-analysis}). Fig. \ref{fig:cavity_gOM}a and Fig. \ref{fig:cavity_gOM}b show the simulated profiles of these modes. The masked implantation process embeds SiV centers within a 62.5 nm x 62.5 nm area (green square shown in Fig. \ref{fig:cavity_gOM}a and Fig \ref{fig:cavity_gOM}b). The TE-like mode has a simulated wavelength of 736 nm and a quality factor of 3.2 $\times$ 10$^4$. The breathing mode has a simulated frequency of 12.29 GHz and a quality factor of 2.4 $\times$ 10$^5$. While the measured optical Q-factor is similar to the simulated quality factor, the measured mechanical Q-factor is much lower than the simulated value.

Finally, Fig. \ref{fig:cavity_gOM}c shows the optomechanical coupling rate of the breathing mode (orange dot) as well as those of all simulated mechanical modes from 5 - 28 GHz, which covers the range of data measured as part of the spin T$_1$ spectrum in Fig. 4. With the exception of modes around 12 GHz, nearly all modes display negligible optomechanical coupling to the TE-like mode due to the strict symmetry requirements for such coupling \cite{chan_optimized_2012}.

\section{Determination of Extrinsic Optical Cavity Loss Rate}\label{extrinsic-loss}
\begin{figure}[h!]
\centering
\includegraphics[width=\textwidth]{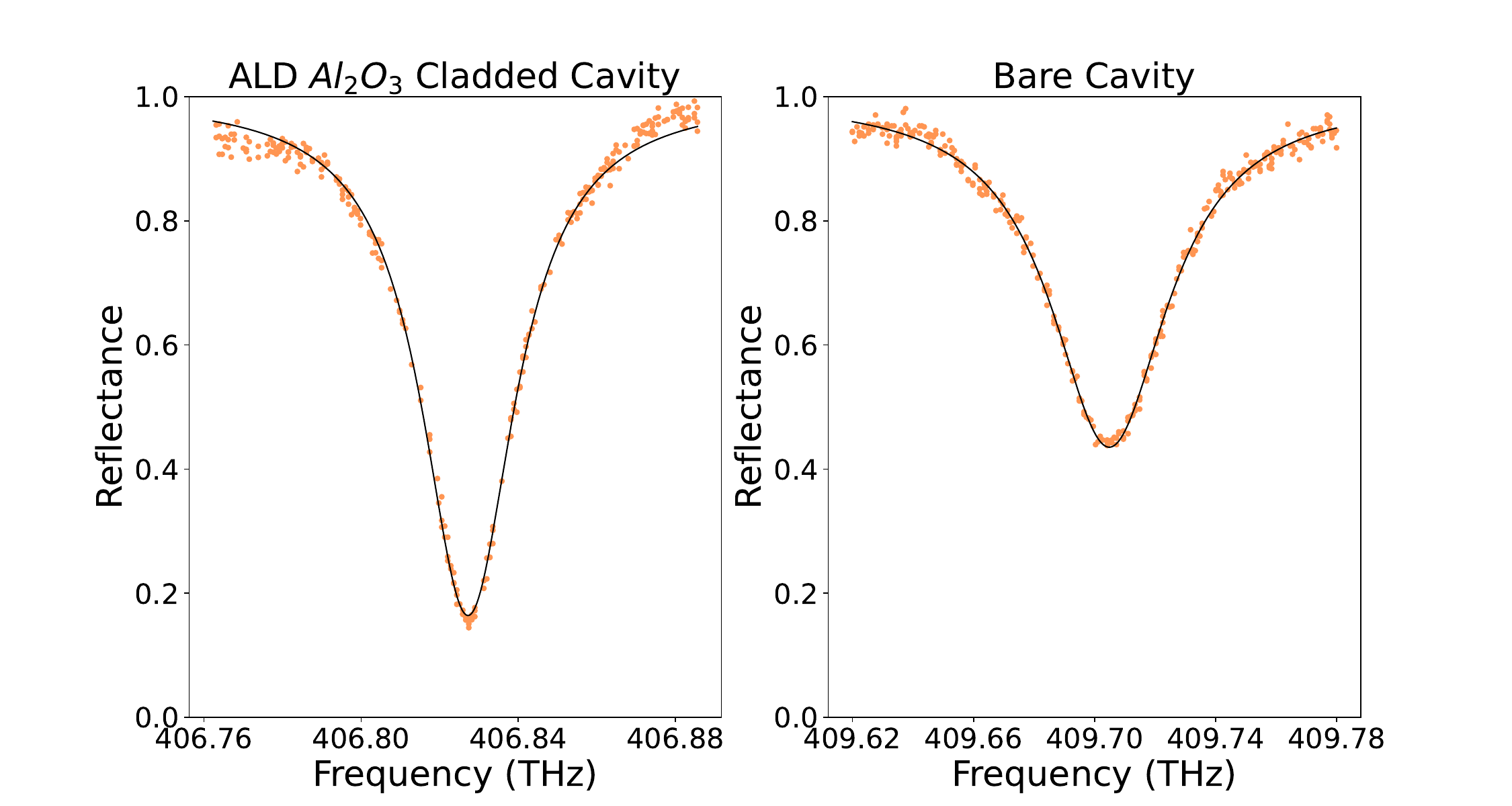}
\caption{Optical reflectance measurement of the bare optical cavity and the cleaned and ALD alumina cladded cavity. The total optical decay rate is $\sim20$ GHz for the bare cavity and $\sim15$ GHz for the ALD alumina clad cavity.}
\label{fig:kappa_extraction}
\end{figure}

It was observed that the quality factor and depth of resonance dip changed after the device was cleaned and cladded in ALD alumina. We suspect, but do not know, that the cleaning was the step responsible for the slight $Q_o$ increase. Fits to both spectra assuming an undercoupled cavity yields $\kappa_e \approx 4$ GHz. Fits assuming an overcoupled system do not yield consistent results for $\kappa_e$ leading us to believe that our devices are undercoupled with $\kappa_i\approx11$ GHz and $\kappa_e \approx 4$ GHz for the ALD alumina-cladded case. This is the best measurement of $\kappa_e$ that we can make in the absence of cavity optical phase response measurements, which are difficult due to the relatively large optical linewidths of our device relative to the available bandwidth of visible-wavelength electro-optic modulators and photodetectors.

\section{Power Dependent Mechanical Linewidth}\label{mechanical-linewidth}
\begin{figure}[h!]
\centering
\includegraphics[width=\textwidth]{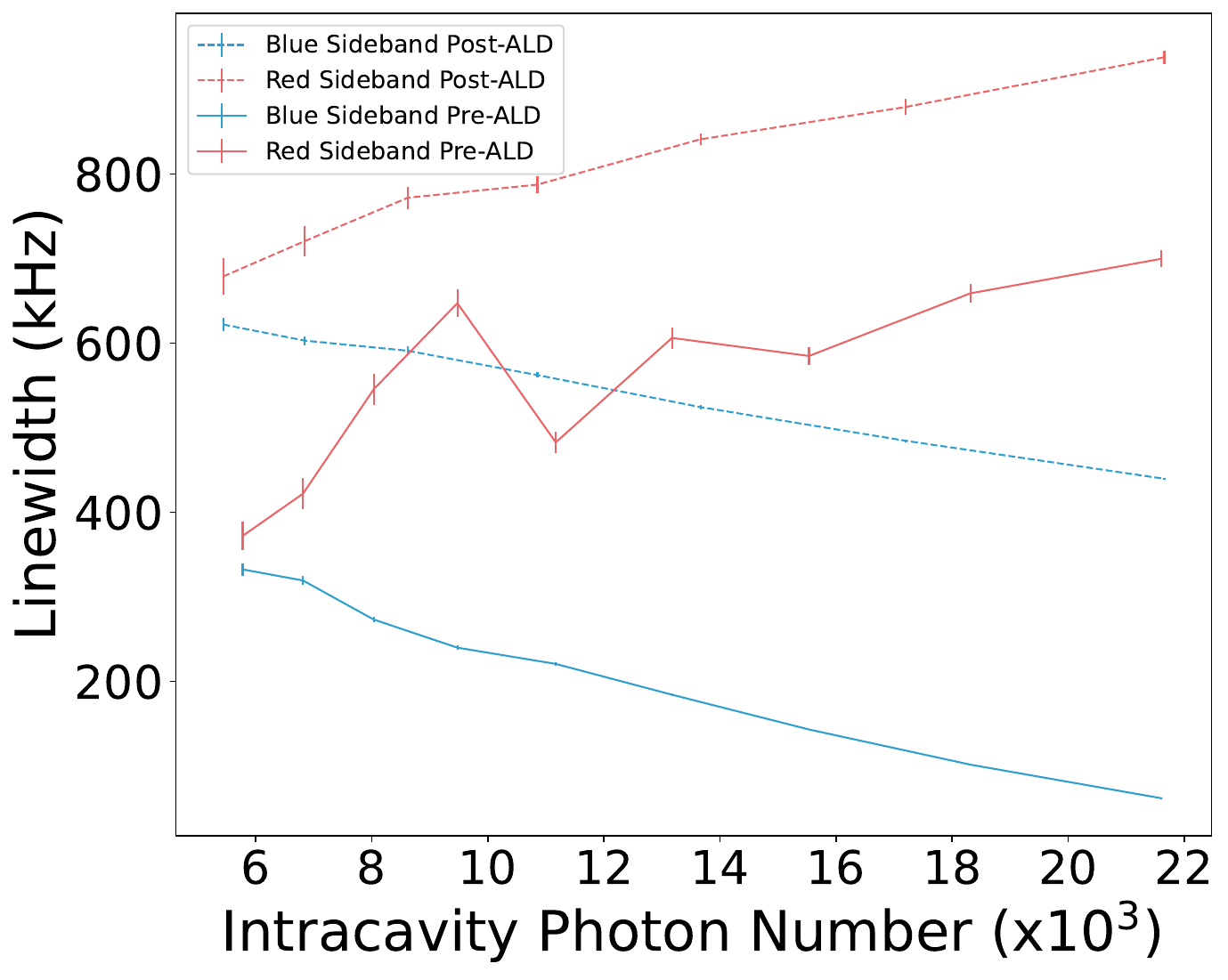}
\caption{Power dependent mechanical mode linewidth with laser parked on the red (blue) sideband for the red (blue) trace before (after) ALD alumina cladding is show with the solid (dashed) lines. Extrapolated intersection of the red/blue sideband curves yield an intrinsic mechanical linewidth of $\sim350$ kHz ($\sim650$ kHz) before (after) ALD alumina cladding.}
\label{fig:power_dependence_optomechanics}
\end{figure}

With sufficiently large input power, the effects of optomechanical backaction can become significant relative to the intrinsic mechanical linewidth of the mechanical mode. Our devices are in the sideband unresolved regime where $\omega_m < \kappa$, so we use an optical probe detuning of $\Delta_o = \pm (\kappa_o/2)/(2\pi)$ to measure the blue (red) sideband optical power dependence of the mechanical mode linewdith. On the blue sideband, optomechanical backaction counteracts the intrinsic damping mechanisms thereby narrowing the mechanical linewidth towards zero (the phonon lasing threshold), whereas on the red sideband, the optomechanical backaction further damps the mechanical mode. The slope of the curves in Fig. \ref{fig:power_dependence_optomechanics} are related to the optomechanical coupling rate, while the y-intercept is the intrinsic mechanical dissipation rate. The solid traces in the lower part of the plot correspond to the bare device, while the dashed traces in the upper part of the plot correspond to the same device after cladding in ALD alumina.

An increase in intrinsic linewidth from $\sim$\SI{350}{\kilo\hertz} to $\sim$\SI{650}{\kilo\hertz} is observed following the cladding of the OMC in alumina. Both of these values are two to three orders of magnitude narrower than the corresponding $\sim$\SI{12}{\giga\hertz} frequency feature observed in the spin decay spectrum measurement in Fig. 3. We attribute this difference to excess mechanical dissipation introduced by the gas deposition procedure required to achieve the optical resonance condition between the TE-like optical mode and the SiV center's ZPL.

\section{Calibration of Spin Transition Frequency}\label{calibration}
\begin{figure}[h!]
\centering
\includegraphics[width=\textwidth]{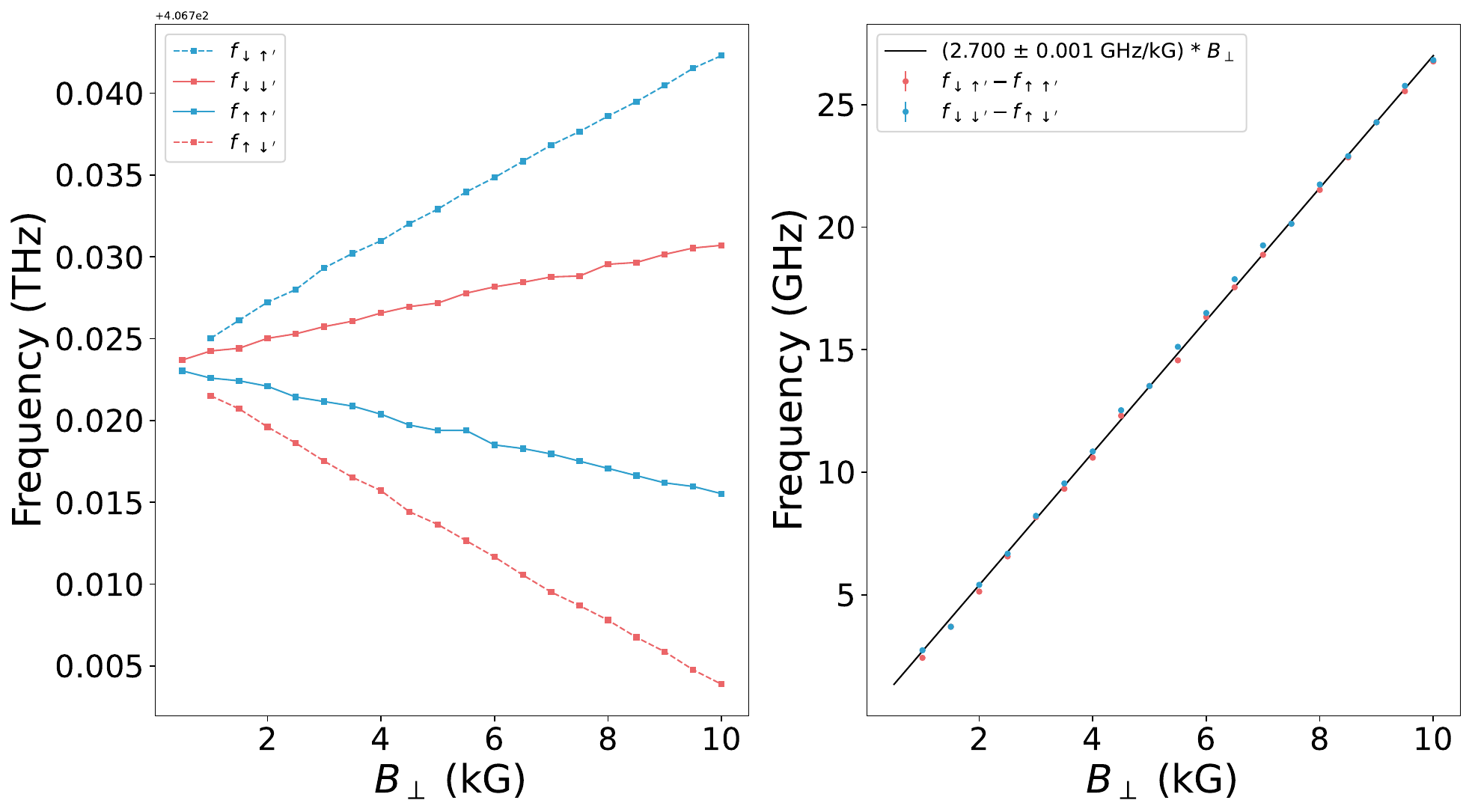}
\caption{Calibration of the spin transition frequency versus applied magnetic field. (a) For any given magnetic field orientation, (in this case perpendicular to the SiV center's symmetry axis), the spin-selective optical transitions shift linearly as a function of the applied field magnitude. (b) The spin transition frequency can be estimated from $f_{\downarrow\uparrow\prime} - f_{\uparrow\uparrow\prime}$ and from $f_{\downarrow\downarrow\prime} - f_{\uparrow\downarrow\prime}$. A linear fit with zero y-intercept to these estimates yields a conversion factor between the magnetic field magnitude and the spin transition frequency (in this case, 2.7 GHz/kG).}
\label{fig:spin_calibration}
\end{figure}

The spin-selective optical transition frequencies depicted in Fig. 2a shift linearly as a function of the magnitude of the applied magnetic field. Specifically, both $f_{\downarrow\uparrow\prime} - f_{\uparrow\uparrow\prime}$ and $f_{\downarrow\downarrow\prime} - f_{\uparrow\downarrow\prime}$ allow us to estimate $\omega_s/(2\pi)$. Here $f_{\uparrow\uparrow\prime}$ and $f_{\downarrow\downarrow\prime}$ are as shown in Fig. 1c. $f_{\downarrow\uparrow\prime}$ is the optical transition between the $\ket{\downarrow}$ and the $\ket{\uparrow\prime}$ state (the leftmost line in Fig. 3a) and $f_{\uparrow\downarrow\prime}$ is the optical transition between the $\ket{\uparrow}$ and the $\ket{\downarrow\prime}$ state (the rightmost line in Fig. 3a). The optical transitions have a maximum Purcell enhanced linewidth of \SI{3.5}{\giga\hertz}, which practically eliminates the effects of spectral diffusion on our experiment. For any given magnetic field orientation, a linear fit to the estimated spin transition frequency at several calibration field magnitudes yields a conversion constant between applied field magnitude and spin transition frequency. In the case of a perpendicular field shown in Fig. \ref{fig:spin_calibration}, we get a conversion constant of $2.700 \pm 0.001$ GHz/kG. We note that using the difference in peak frequencies of these broad optical lines is not a direct measurement of the narrow spin transition frequency. While the statistical uncertainty in this estimator is very small, reflecting its high precision, this does not account for systematic errors like differential AC stark shifts between the various optical line measurements which limit its accuracy. Here, we use these calibrations as a coarse spin-transition frequency measurement and then use sharp identifiable features in the observed spectra to line up and compare measurements under different conditions. Observation of the $\sim$12 GHz mode (whose frequency is known to high precision from optomechanical measurements) then offers a useful reference to identify the frequency of other features with high confidence.

For our spin decay spectroscopy measurements, we use the $f_{\uparrow\uparrow\prime}$  and $f_{\downarrow\downarrow\prime}$ transitions because of their large photon scattering rate and comparatively small movement as a function of the applied magnetic field magnitude. The value of the conversion constant depends on the angle of the applied magnetic field and is calibrated at every angle for the angle-dependent measurements in Fig. 4. In particular, for a given field magnitude, a magnetic field oriented perpendicular to the SiV center's symmetry axis will produce the minimum spin transition frequency, whereas a field aligned with the symmetry axis will produce the maximum spin transition frequency. The value of the conversion constant also depends on the static crystal strain experienced by the SiV center, which can be inferred from the ground state orbital splitting \cite{meesala_strain_2018}. It was observed to change after ALD alumina cladding (see SI section \ref{static-strain-measurement}), and to a lesser extent after repeated gas tuning. Therefore, this calibration curve is retaken after any material deposition performed to tune the optical resonance.

\section{Raw Data for Spin Relaxation Measurements}\label{raw-data}

\begin{figure}[h]
\centering
\includegraphics[width=\textwidth]{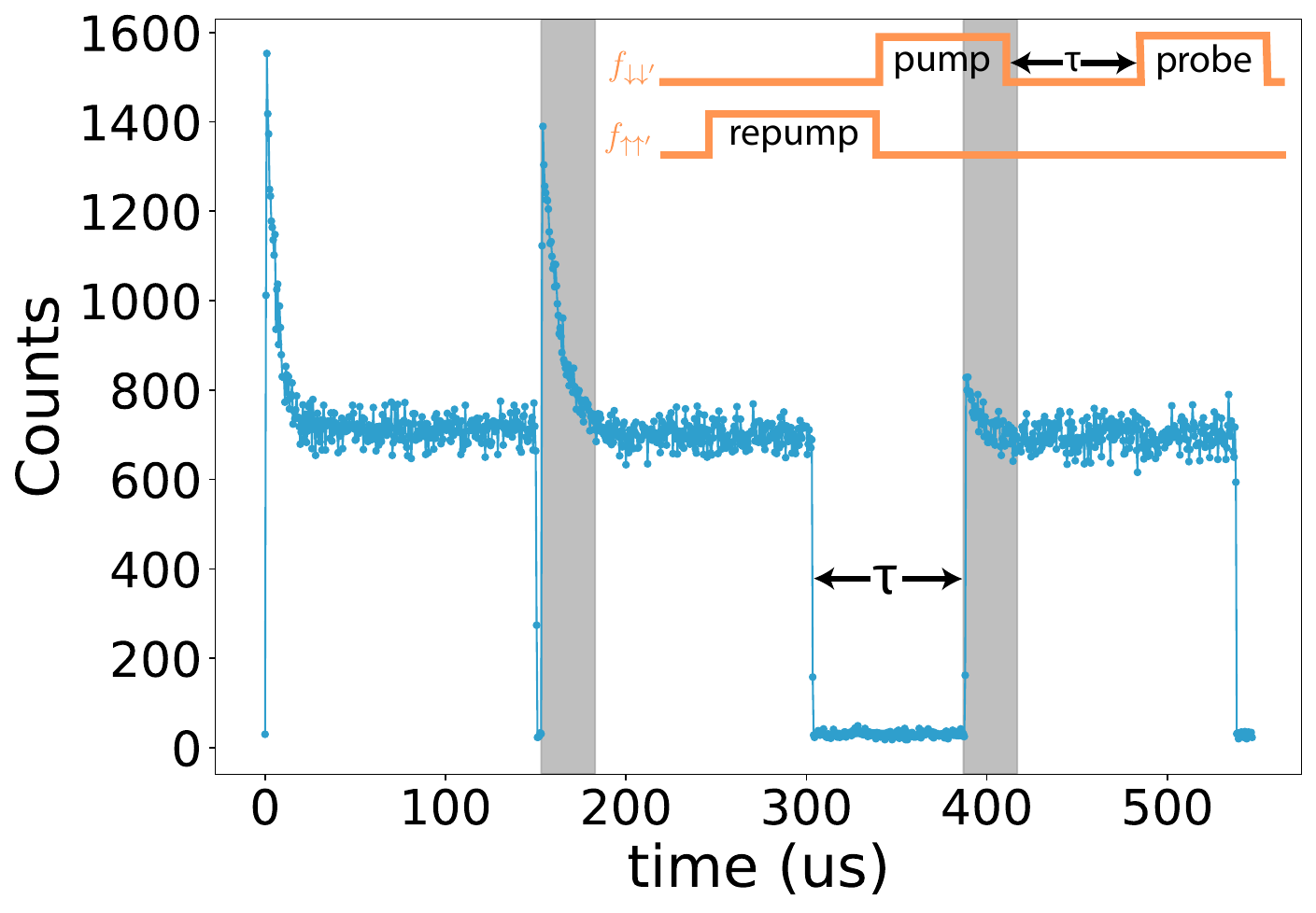}
\caption{An example histogram of photon detection times collected from the inset optical pulse sequence and a particular value of $\tau$.}
\label{fig:histogram}
\end{figure}

An example histogram of the photon counts measured by the APD connected to the time-tagger for a particular time delay $\tau$ of the optical pulse sequence is shown in Fig. \ref{fig:histogram}. The initialization peak seen at the beginning of each pulse corresponds to a short time of increased reflection from the device as the spin is pumped to the opposite state (shaded). The repump pulse at the beginning of the sequence, which is resonant with the f$_{\uparrow\uparrow\prime}$ transition, is intended to ensure that the spin population begins predominantly in the $\ket{\downarrow}$ state. Following this repump pulse, the pump and probe pulses are resonant with the f$_{\downarrow\downarrow\prime}$ transition. The pump pulse then pumps the spin population into the $\ket{\uparrow}$ state and produces an initial peak fluorescence corresponding to maximum population in the $\ket{\uparrow}$ state. After the variable wait time $\tau$, the height of the initial fluorescence signal on the probe pulse relative to the height of the initialization peak on the pump pulse is a measurement of the fraction of the population that has relaxed into the $\ket{\downarrow}$ state. As $\tau$ is increased beyond the characteristic spin T\textsubscript{1} time, the SiV center’s spin population decays towards its thermal equilibrium value. The f$_{\uparrow\uparrow\prime}$ and f$_{\downarrow\downarrow\prime}$ transitions are nominally spin-conserving, but in a magnetic field not aligned with the SiV center symmetry axis, this spin selection rule is relaxed and the transitions can be used to polarize the spin. In some of our spin relaxation measurements, we found that the repump pulse does not completely polarize the spin population with the maximum fidelity, however the timescale of the spin relaxation, which is the quantity of interest in our measurements, is unaffected.

\section{Temperature-Dependent Spin Decay}\label{temperature-dependence}
\begin{figure}[h!]
\centering
\includegraphics[width=\textwidth]{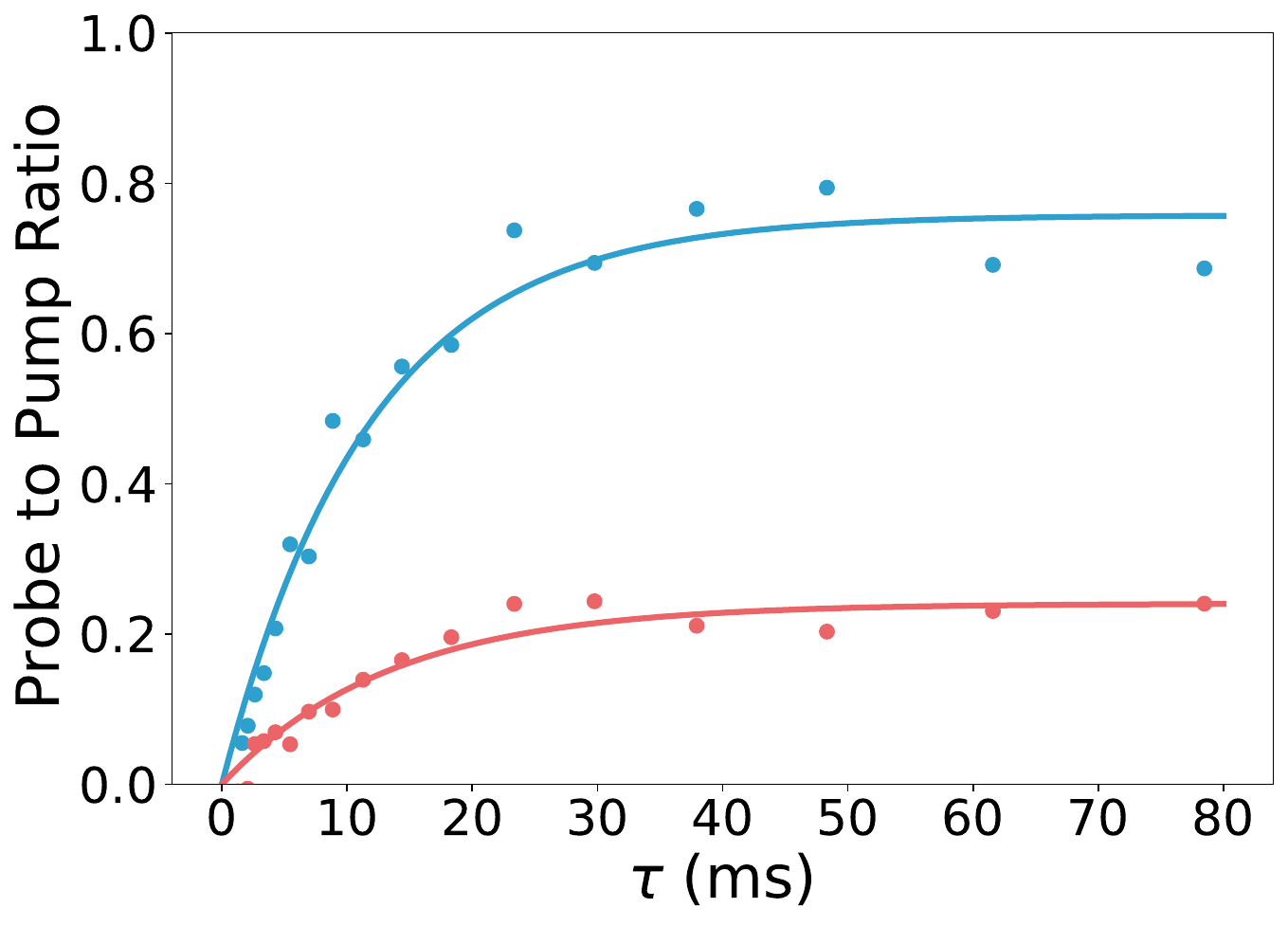}
\caption{Example spin decay curves measured with a 1 kG magnetic field with $\theta\approx55$\SI{}{\degree} (corresponding to a spin transition frequency of 2.8 GHz). Blue (red) curves measured with the pump/probe laser parked on the $f_{\uparrow\uparrow\prime}$ ($f_{\downarrow\downarrow\prime}$) transition.}
\label{fig:base_temperature_measurement}
\end{figure}

\begin{figure}[h!]
\centering
\includegraphics[width=\textwidth]{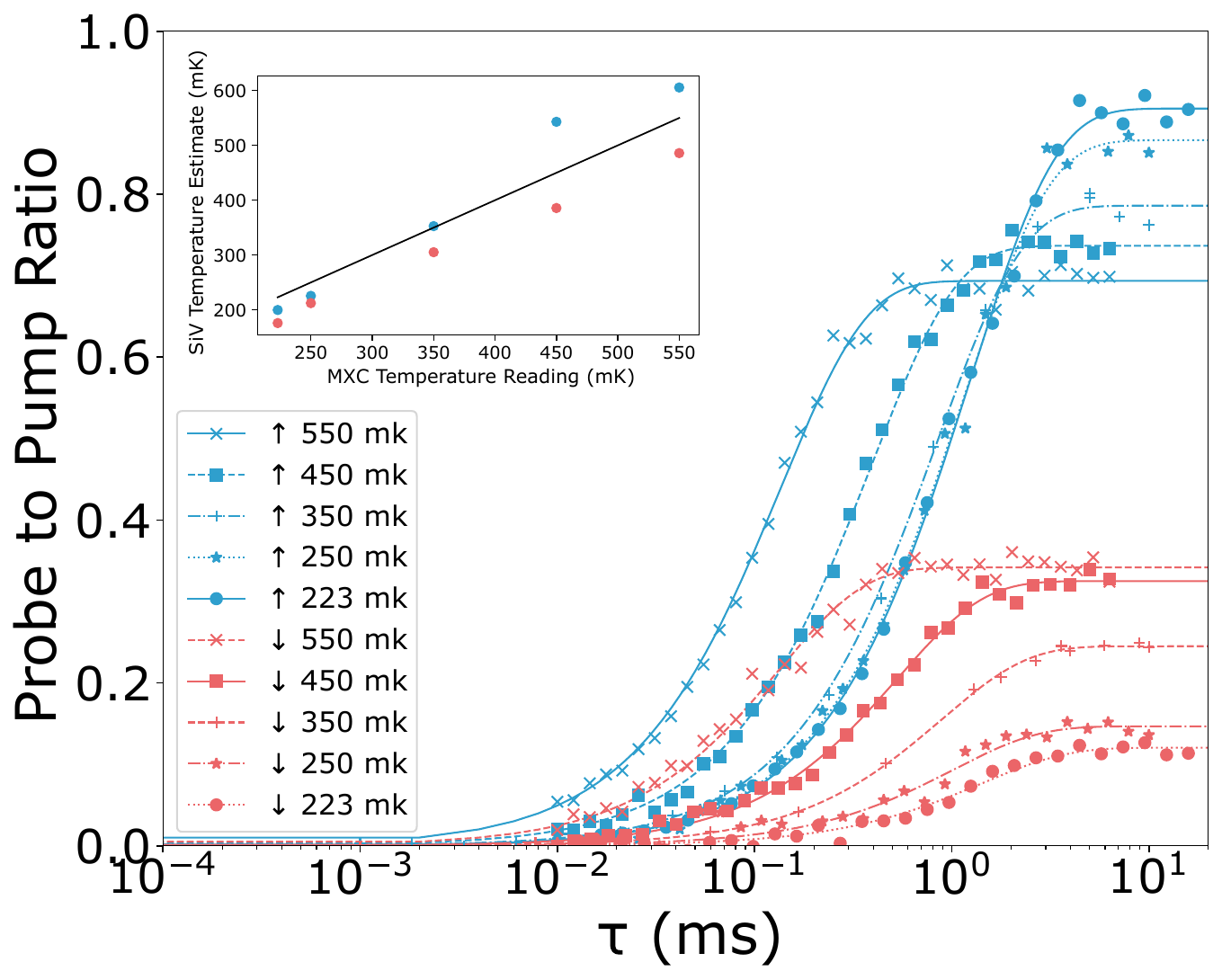}
\caption{Temperature dependent spin decay curves. With an applied magnetic field of 3 kG aligned with the laboratory z-axis (corresponding to a spin transition frequency of 8.3 GHz), the spin decay is measured at five different elevated temperatures. Measurements are performed on both probe transitions while applying controlled heat to the mixing plate and keeping the optical probe powers constant. Using the steady state population, an SiV temperature estimate was extracted. The inset shows a plot of this estimated SiV temperature against the temperature recorded on the MXC thermometer of the dilution fridge.}
\label{fig:spin_temperature}
\end{figure}

We performed thermometry measurements by probing the SiV center spin relaxation with low input laser powers on the order of 1 pW. The measurement in Fig. \ref{fig:base_temperature_measurement} shows that intialization in the spin up state (blue data in Fig. \ref{fig:base_temperature_measurement}) produces a greater contrast in the spin relaxation curve at long time delays (relative to $1/\gamma_s$) when compared with initialization in the spin down state (red data in Fig. \ref{fig:base_temperature_measurement}). The data is taken at a magnetic field magnitude of 1 kG and orientation of $\theta\approx55$\SI{}{\degree}. The characteristic exponential timescale of spin relaxation for both cases is similar. At long time delays, the spin will relax to a thermal state. Based on the temperature of the SiV's environment, T and the energy difference between the spin qubit levels, $\hbar\omega_s$ the steady-state population in the spin up and down states is given by the relations:

\begin{equation}
    p_{\uparrow}(\tau\gg T_1)=\frac{1}{e^{\frac{\hbar\omega_s}{k_B T}}+1}
\end{equation}

\begin{equation}
    p_{\downarrow}(\tau\gg T_1)=1-\frac{1}{e^{\frac{\hbar\omega_s}{k_B T}}+1}
\end{equation}

Our measurements yield an estimated SiV temperature of $\sim150$ mK in the steady state. This is elevated above the thermometer readings on the copper sample mount (65 mK) and the mixing plate thermometer (44 mK), likely due to residual optical probe heating \cite{knall_efficient_2022}. Nonetheless, at this temperature of $\sim150$ mK, the 12 GHz mechanical breathing mode of our device is expected to be close to the ground state with an expected thermal phonon occupancy of 0.02 quanta. This temperature is also significantly below the temperature for $>99\%$ population in the lower orbital branch of the SiV ground state, calculated to be $\sim$\SI{885}{\milli\kelvin}, given $\Delta_{gs} = 85$ GHz (see SI section \ref{static-strain-measurement}). Around this temperature, the phonon population resonant with the 85 GHz ground state orbital transition can act as an additional channel for spin relaxation via an Orbach process which can dominate the direct spin-relaxation pathway \cite{orbach_spin-lattice_1997, meesala_strain_2018}.

Fig. \ref{fig:spin_temperature} shows the temperature-dependent spin relaxation data measured by applying a controlled amount of heat to the mixing plate of the dilution fridge. As the mixing plate temperature is raised, the spin qubit is found to decay faster with a steady state ($\tau\gg1/\gamma_s$) population value tending towards 0.5 which corresponds to a completely mixed spin state. This behavior is expected at higher temperatures relative to the spin transition frequency. The SiV temperature estimate for each measurement can be extracted from the long-time population saturation value and is shown inset to Fig. \ref{fig:spin_temperature}. In the temperature range probed in these measurements, the SiV temperature estimate from this method is in good agreement with the temperature reading on the mixing plate thermometer.

\section{Simulation of Spin-Phonon Coupling to the Mechanical Breathing Mode}
\label{breathing_mode_simulation}
\begin{figure}[h!]
    \centering
    \includegraphics[width=\linewidth]{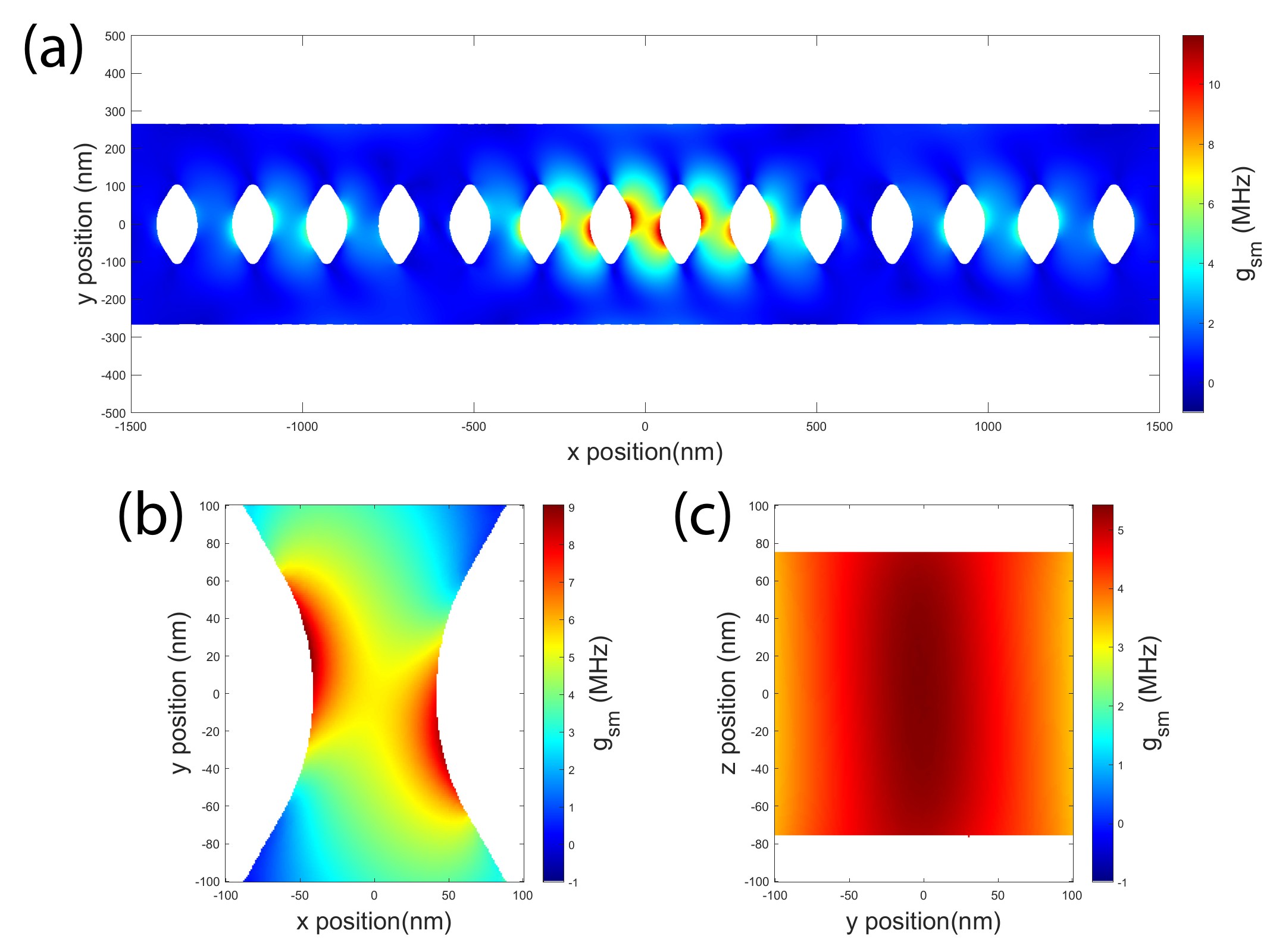}
    \caption{(a) XY slice plot of simulated $g_{sm}$ in the cavity region. Large $g_{sm}$ can be realized where strain is concentrated near the cavity center.(b) XY slice plot of simulated $g_{sm}$ for the implantation aperture region. $g_{sm}$ is sensitive to the in-plane location, with maximal $g_{sm}$ found at the hole edges where strain is maximized.  (c) YZ slice plot of simulated $g_{sm}$ for the aperture region. The SiV centers are nominally implanted at the beam center (z=0), but $g_{sm}$ is insensitive to depth.}
    \label{fig:breathing_mode}
\end{figure}

The strain profile of the breathing mode is used to calculate the strength of the spin-phonon coupling $g_{sm}$ for an ideal zero static-strain SiV. Using perturbation theory \cite{meesala_strain_2018,chia_quantum_2021} the vacuum spin-phonon coupling rate $g_{sm}$ is given by:

\begin{equation}
    g_{sm} = \frac{2\gamma_sB_x}{\lambda_{SO,GS}} \sqrt{\beta^2+\gamma^2}
\end{equation}

where 

\begin{align}
    \beta = d(\epsilon_{XX} - \epsilon_{YY}) + f \epsilon_{XZ}\\ \gamma = - 2d\epsilon_{XY}+f\epsilon_{YZ}
\end{align}

and the components of the static strain tensor $\epsilon$ are in the coordinate frame of the SiV center.

With the exception of Fig. 4b, Extended Data Fig. 3, Extended Data Fig. 4, and Extended Data Fig. 5, the data shown in this paper was taken using a magnetic field aligned to the z- axis in laboratory frame ($\theta \approx55$\SI{}{\degree}). In this case, the magnetic field component perpendicular to the SiV center's high-symmetry axis $B_X$ is related to the spin transition frequency by $\omega_s$ by equation $\gamma_s B_X \approx \frac{\omega_s}{\sqrt{2}}$, on resonance $\omega_s=\omega_m$. The spin-mechanical coupling rate can then be expressed as:

\begin{equation}g_{sm} \approx \frac{\sqrt2 \omega_m}{\lambda_{SO}} \sqrt{(d(\epsilon_{XX}-\epsilon_{YY}) + f\epsilon_{YZ})^2  + (-2d\epsilon_{XY} + f\epsilon_{XZ})^2}
\end{equation}

The SiV center's spin transition is sensitive to the strain transverse to its symmetry axis, precisely at the SiV center’s location in the cavity. Large $g_{sm}$ can be achieved  in areas of maximal strain as seen in Fig. \ref{fig:breathing_mode} for the mechanical breathing mode of our OMC devices. Fig. \ref{fig:breathing_mode}c shows that $g_{sm}$ is largely insensitive to depth, but highly sensitive to in-plane location. Simulations suggest that under ideal conditions of low crystal strain (strain $\ll \lambda_{so}=46$ GHz in the orbital ground state) and a magnetic field orientation perpendicular to the high-symmetry axis of the SiV \cite{meesala_strain_2018}, $g_{sm}$ can reach a maximum value of 9 MHz. This is roughly a factor of 30 larger than the experimentally observed $g_{sm}$. This difference is well explained by the non-ideal conditions in our OMC devices, in particular the non-zero static crystal strain (see SI section \ref{static-strain-measurement}) and expected error in the SiV center's position with respect to the strain coupling maximum.

\section{Measurement of SiV Static Strain Environment}\label{static-strain-measurement}

\begin{figure}[h!]
\centering
\includegraphics[width=\textwidth]{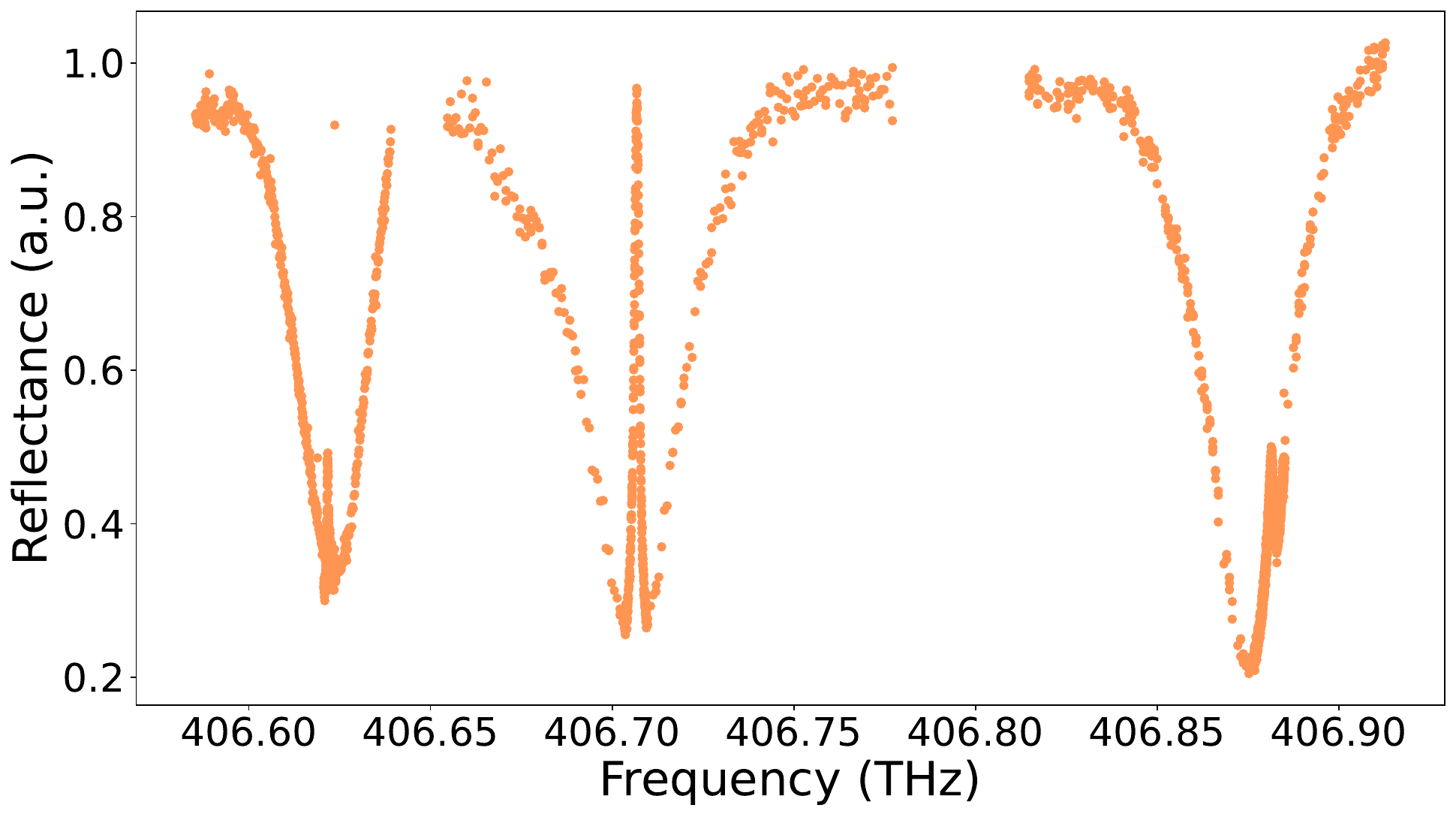}
\caption{Optical reflection spectrum of the SiV cavity system with the optical cavity (broader feature) gas tuned into resonance with the D (left), C (middle), and B (right) optical transitions in the SiV fine structure. From the measured optical transition frequencies we infer a ground state splitting $\Delta_{gs}=85$ GHz and an excited state splitting $\Delta_{es}=260$ GHz corresponding to a moderately strained SiV center.}
\label{fig:strain_extraction}
\end{figure}

Using the gas tuning capabilities of our experiment, it is possible to tune the optical resonance of the TE-like optical mode of our devices into resonance with each of the lines in the fine structure of an optically coupled SiV center's ZPL \cite{hepp_electronic_2014}. These measurements are performed at zero magnetic field. At $\sim$4K, a distinct narrow resonance within the optical reflection spectrum is observed at the frequency resonant with the atomic line \cite{waks_dipole_2006, sipahigil_integrated_2016}. Knowledge of the frequency splittings of an SiV center's fine structure is useful because it can be used as a measurement of the static crystal strain experienced by the SiV center. The data in Fig. \ref{fig:strain_extraction} was taken after the device was clad in ALD alumina making it impossible to reach the bluest transition (the A-line). However, the observation of three lines is sufficient to measure the ground and excited state splittings ($\Delta_{gs}=85$ GHz and $\Delta_{es}=260$ GHz respectively). Note that at millikelvin temperatures where the ground state phonon transition is frozen out, the B and D line go dark due to a lack of population in the upper branch of the ground state, making them unsuitable for measurements in this temperature range.

\section{Simulation of Spin-Phonon Coupling to the Broadband Acoustic Density of States}\label{gamma_simulation}
\begin{figure}[h!]
\centering
\includegraphics[width=.8\textwidth]{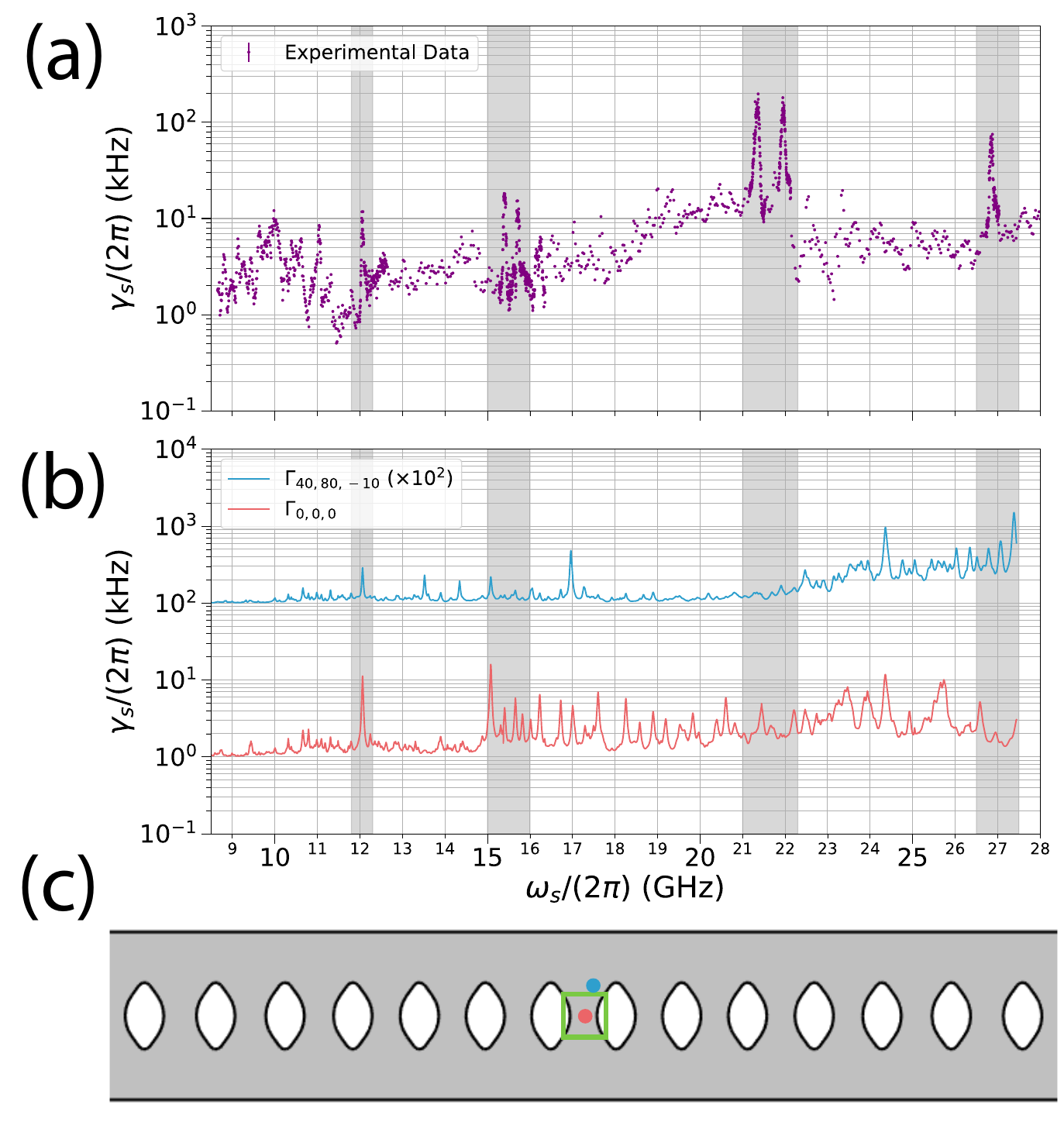}
\caption{(a) Broadband spin decay spectrum for a $\theta\approx$ \SI{55}{\degree} magnetic field, also shown in Fig. 4. Shaded regions indicate observed resonances of increased Purcell enhancement, attributed to localized mechanical modes. (b) Simulated spin decay spectrum $\Gamma_{x,y,z}$ for different choices of SiV-center location relative to the cavity center (in nm). Note that some experimental features are reproduced (namely Purcell enhancement due to the $\sim12$ GHz breathing mode) for both locations, but qualitatively distinct features are observed at other frequencies. (c) Cavity diagram indicating the chosen points for simulation, the implantation aperture is shown in green.}
\label{fig:gamma_simulation}
\end{figure}

While our design work focused on the mechanical breathing mode, the coupling of the SiV center to mechanical modes in the 8.5-28 GHz frequency range was measured and simulated as seen in Fig. \ref{fig:gamma_simulation}. To model the phonon-induced decay of the SiV center as a function of the spin transition frequency, we sum the contributions from all of these modes:

\begin{equation}
    \Gamma(\omega) = 4 \sum_q \frac{g_q^2\kappa_q}{\kappa_q^2+4(\omega_q - \omega)^2}
\end{equation}

To account for the fact that the experimentally observed quality factor ($Q_q = \frac{\omega_q}{\kappa_q}$) is lower than what is simulated, we include a phenomenological reduction in Q so that $Q_q' = (Q_{sim}^{-1} + Q_{damp}^{-1})^{-1}$, where $Q_{damp} = 350$ based on the experimentally observed linewidth broadening of the breathing mode. We also account for the static-strain of the SiV as measured in section \ref{static-strain-measurement}, which leads to a reduction in $g_q$.

The highly local nature of spin-phonon coupling poses a modeling challenge due to the uncertain position of the SiV center within the implantation aperture, which is $50$ nm wide in both the x and y directions. As seen in Fig. \ref{fig:gamma_simulation}, even a few nanometer change of the in-plane location of the SiV center results in qualitatively distinct spin decay spectra. While this sensitivity could, in principle, be used to determine the precise SiV location, in practice, the complexity of the spectrum and the difficulty of modeling real-world imperfections—such as the aforementioned Q-factor degradation from ALD alumina and deposited gas (as shown in Fig. 3b and Extended Data Fig. 3) and device roughness and non-uniformity resulting from the quasi-isotropic etching method, limit the accuracy of such simulations. Nonetheless, many of the experimentally observed features are seen in our simulations. In particular, a resonance in the spin decay rate due to the 12 GHz breathing mode of the device can be observed consistently. The agreement of these simulations with both the spin-decay spectroscopy and optomechanical spectroscopy measurements on the 12 GHz breathing mode support the validity of this simulation approach.

\printbibliography[title={Supplementary Information References}]
\end{refsection}
